\documentclass[conference]{IEEEtran}
\IEEEoverridecommandlockouts
\usepackage{amsmath,amssymb,amsfonts}
\usepackage{adjustbox}
\usepackage{algorithm}
\usepackage{algpseudocode}
\usepackage{booktabs}
\usepackage{caption}
\usepackage{color}
\usepackage{colortbl}
\usepackage{cite}
\usepackage{float}
\usepackage{graphicx}
\usepackage{hhline}
\usepackage{hyperref}
\usepackage[utf8]{inputenc}
\usepackage{listings}
\usepackage{textcomp}
\usepackage{xcolor}
\usepackage{marvosym}
\usepackage{multirow}
\usepackage{tabularx}
\usepackage{siunitx}
\usepackage{subcaption}
\usepackage{threeparttable}
\usepackage{tikz}
\usepackage{url}
\usetikzlibrary{positioning, arrows.meta}

\definecolor{moduleblue}{RGB}{31,119,180}
\definecolor{datagreen}{RGB}{44,160,44}
\definecolor{processorange}{RGB}{255,127,14}
\definecolor{confred}{RGB}{214,39,40}
\definecolor{tablegray}{RGB}{245,245,245}

\begin{document}

\title{MultiRAG: A Knowledge-guided Framework for Mitigating Hallucination in Multi-source Retrieval Augmented Generation
}
\DeclareRobustCommand*{\IEEEauthorrefmark}[1]{%
  \raisebox{0pt}[0pt][0pt]{\textsuperscript{\footnotesize #1}}%
}

% \author{\IEEEauthorblockN{Anonymous Authors}}

\author{
	\IEEEauthorblockN{
		Wenlong Wu \IEEEauthorrefmark{1},
        Haofen Wang \IEEEauthorrefmark{2}\thanks{\textit{Wenlong Wu and Haofen Wang contributed equally to this work.}}, Bohan Li \IEEEauthorrefmark{1,3,4}\textsuperscript{\Letter}\thanks{\textit{Bohan Li is the corresponding author.}},
        Peixuan Huang \IEEEauthorrefmark{1}, Xinzhe Zhao \IEEEauthorrefmark{1}
		and Lei Liang\IEEEauthorrefmark{5}} 
\IEEEauthorblockA{
    \IEEEauthorrefmark{1}College of Artificial Intelligence, Nanjing University of Aeronautics and Astronautics,\\
    Key Laboratory of Brain-Machine Intelligence Technology, Ministry of Education\\
    }
    
        % \IEEEauthorrefmark{2}Key Laboratory of Brain-Machine Intelligence Technology, Ministry of Education\\
\IEEEauthorblockA{
    \IEEEauthorrefmark{2}College of Design \& Innovation, Tongji University
}
\IEEEauthorblockA{
    \IEEEauthorrefmark{3}Key Laboratory of Intelligent Decision and Digital Operation, Ministry of Industry and Information Technology
}
\IEEEauthorblockA{
    \IEEEauthorrefmark{4}Collaborative Innovation Center of Novel Software Technology and Industrialization
}
    \IEEEauthorblockA{\IEEEauthorrefmark{5}Ant Group Knowledge Graph Team\\ 
    Email: \{wuwenlong, bhli, peixuanh, xinzhe\_zhao\}@nuaa.edu.cn\\
    carter.whfcarter@gmail.com, leywar.liang@antgroup.com}
}

\maketitle

% 检索增强生成已经被用来解决LLMs的幻觉问题。但是由于检索源的质量参差不齐，以及检索与LLM交互的过程都会引入新的噪声，从而导致幻觉无法完全避免。在面向多源异构数据检索时，幻觉现象会更加严重。相较于单源数据的检索，多源数据检索会产生额外的两个挑战，包括源数据分布稀疏，导致跨源逻辑关系难以捕捉的问题;以及源间数据不一致，造成的信息冲突问题。鉴于此，本文提出一种新颖的方法MultiRAG，用于消除多源知识增强生成中的幻觉现象。我们的工作主要包含以下两个方面。 首先，在知识构建模块，我们引入多源线图，实现异构知识源间逻辑关系的高效聚合，以解决数据稀疏分布带来的问题。第二，在检索模块，我们采用多层级置信度计算方法，在多源线图上进行图级别和节点级别的置信度计算，剔除不可靠节点，以缓解源间数据不一致带来的幻觉问题。最后，我们在四个开源多域查询数据集和两个多跳知识图谱问答数据集上进行广泛的实验，验证了MultiRAG在面向复杂多源异构数据检索时的鲁棒性和高效性。

\begin{abstract} 

Retrieval Augmented Generation (RAG) has emerged as a promising solution to address hallucination issues in Large Language Models (LLMs). However, the integration of multiple retrieval sources, while potentially more informative, introduces new challenges that can paradoxically exacerbate hallucination problems. These challenges manifest primarily in two aspects: the sparse distribution of multi-source data that hinders the capture of logical relationships and the inherent inconsistencies among different sources that lead to information conflicts. To address these challenges, we propose MultiRAG, a novel framework designed to mitigate hallucination in multi-source retrieval-augmented generation through knowledge-guided approaches. Our framework introduces two key innovations: (1) a knowledge construction module that employs multi-source line graphs to efficiently aggregate logical relationships across different knowledge sources, effectively addressing the sparse data distribution issue; and (2) a sophisticated retrieval module that implements a multi-level confidence calculation mechanism, performing both graph-level and node-level assessments to identify and eliminate unreliable information nodes, thereby reducing hallucinations caused by inter-source inconsistencies. Extensive experiments on four multi-domain query datasets and two multi-hop QA datasets demonstrate that MultiRAG significantly enhances the reliability and efficiency of knowledge retrieval in complex multi-source scenarios. \textcolor{blue}{Our code is available in \href{https://github.com/wuwenlong123/MultiRAG}{https://github.com/wuwenlong123/MultiRAG}.}
\end{abstract}


% \begin{equation}
%     FusionData =\sum_{d\in D_{stru}}{Ada_{stru}(d)}+\sum_{d\in D_{semi-s}}{Ada_{semi-s}(d)}+\sum_{d\in D_{unstru}}{Ada_{unstru}(d)}
% \end{equation}
\section{Introduction}
% 大型语言模型（LLMs）已经在处理各类自然语言处理任务上取得了引以为傲的成绩，这归因于它们对语言和符号的强大的理解和生成能力[引用]。在知识密集型的检索任务上，检索增强生成（RAG）已然成为一个标准化的解决范式。在先前的工作里\cite{ref31,ref32,ref33,ref34,metarag,ref35}，RAG在解决LLMs的固有知识限制方面取得了重要进展。它通过外部知识库的引入，显著提高了LLMs回答的准确性和忠实度。然而，最近的研究强调了目前仍然存在一个显著的缺点，RAG的检索结果是不完美的，这包括不相关的、误导性的，甚至是恶意的信息\ref[3个]，这最终导致不准确的LLM响应。
    
Large Language Models (LLMs) have achieved remarkable success in handling a variety of natural language processing tasks, attributable to their robust capabilities in understanding and generating language and symbols \cite{ref1}. In knowledge-intensive retrieval tasks, Retrieval Augmented Generation (RAG) has become a standardized solution paradigm\cite{rag1,rag2,gao2023retrieval}. Previous works \cite{ref31,ref32,ref33,ref34,metarag,ref35,2024-40634} have made significant strides in addressing the inherent knowledge limitations of LLMs. By introducing external knowledge bases, it has markedly improved the accuracy and fidelity of LLM responses. However, recent studies have highlighted a significant drawback: the retrieval results of RAG \textcolor{black}{are} imperfect, including irrelevant, misleading, and even malicious information, ultimately leading to inaccurate LLM responses.

% 于是，LLM和KG的协同被提出来实现更高效的信息检索\cite{ref41}。一方面，知识图谱可以高效存储具备固定特征的数据（如时序KG，事理KG等），从而增强LLMs在特定数据上的处理能力\cite{ref51,ref52,ref53,ref54})。另一方面，LLM和KG的协同很好地提高了多跳问答和多文档问答领域的表现，包括检索的可信和可解释性\cite{RoG}。此外，GraphRAG\cite{ref7}、KAG\cite{liang2024kag}等 LLM-KG 协同方法也提供了知识密集型检索任务的最新的解决方案，推动RAG的深度推理能力。

\textcolor{black}{To address these limitations}, the synergy between LLMs and Knowledge Graphs (KGs) has been proposed to achieve more efficient information retrieval \cite{ref41}. On one hand, KG can efficiently store data with fixed characteristics (such as temporal KGs, event KGs, etc.), thereby enhancing the processing capabilities of LLMs on specific data \cite{ref51,ref52,ref53,ref54,li2024leveraging,hu2024decoupling,li2025hycube,li2024hje}. On the other hand, the collaboration between LLMs and KGs has significantly improved performance in \textcolor{black}{multi-hop and multi-document question answering}, including the credibility and interpretability of retrieval \cite{RoG}. Furthermore, LLM-KG collaborative methods have also provided the latest solutions for knowledge-intensive retrieval tasks \cite{graphrag, hipporag, gnnrag, tog, liang2024kag}, propelling the deep reasoning capabilities of RAG.

% 即便如此,现有的框架仍然对真实世界中数据的复杂性欠缺考虑。即便RAG能在一定程度上抑制幻觉的产生，但这些幻觉往往来自于LLMs的内部知识\ref[ref9]。不一致的信息源和不可靠的检索手段\ref[metarag]，仍会致使LLM的检索偏差和幻觉。尤其是在面向大规模多源知识的信息检索任务时，幻觉现象变得尤为突出。研究显示\ref[astute]，约70\%的检索段落不直接包含正确的查询答案，但记录了与答案间接相关的信息，对LLMs造成了误导和理解偏差。
\textcolor{black}{Nevertheless, existing frameworks still fail to account for the complexity of real-world data.} Although RAG can mitigate the generation of hallucinations, these hallucinations often stem from the internal knowledge of LLMs \cite{ref9,wang2024large,gao2025u}. Inconsistent information sources and unreliable retrieval methods can still lead to retrieval biases and hallucinations in LLMs. This issue becomes particularly pronounced when dealing with information retrieval tasks that involve multi-source knowledge, where hallucinations are more prominent. Research \cite{astute} indicates that approximately 70\% of retrieved paragraphs do not directly contain the correct query answers but instead include information indirectly related to the answers, causing misguidance and comprehension bias in LLMs.

% 受益于\cite{metarag}对检索中出现错误的划分，本文列举出多源数据检索中最为常见的三类幻觉:
% 1. 源间数据不一致性：不同数据源之间的数据差异可能导致信息冲突，造成LLMs的幻觉，影响检索结果的准确性。

% 2. 相似数据的冗余：多个数据源间往往存在十分相似但并不具备同样语义的数据，这会给造成极大的检索计算开销，同时带来幻觉现象。

% 3. 推理路径不完整：从不同数据源整合信息以形成全面的推理路径是一项挑战。现有的检索器往往无法捕捉到多个数据源中的完整逻辑关联。
Building upon the categorization of hallucinations in retrieval \cite{metarag}, we outlines the three most common types of hallucinations encountered in multi-source data retrieval:
\begin{enumerate}
    \item \textbf{Inter-source data inconsistency}: Discrepancies between different data sources can lead to conflicting information, causing hallucinations in LLMs.
    \item \textbf{Redundancy of similar data}: There often exists data that is highly similar and semantically equivalent across multiple data sources, which can impose significant computational overhead on retrieval.
    \item \textbf{Incomplete inference paths}: Forming a comprehensive inference path from different data sources is challenging. Existing retrievers often fail to capture the complete logical associations within multiple data sources.
\end{enumerate}

\begin{figure}[t]
    \centering
    \includegraphics[width=\linewidth]{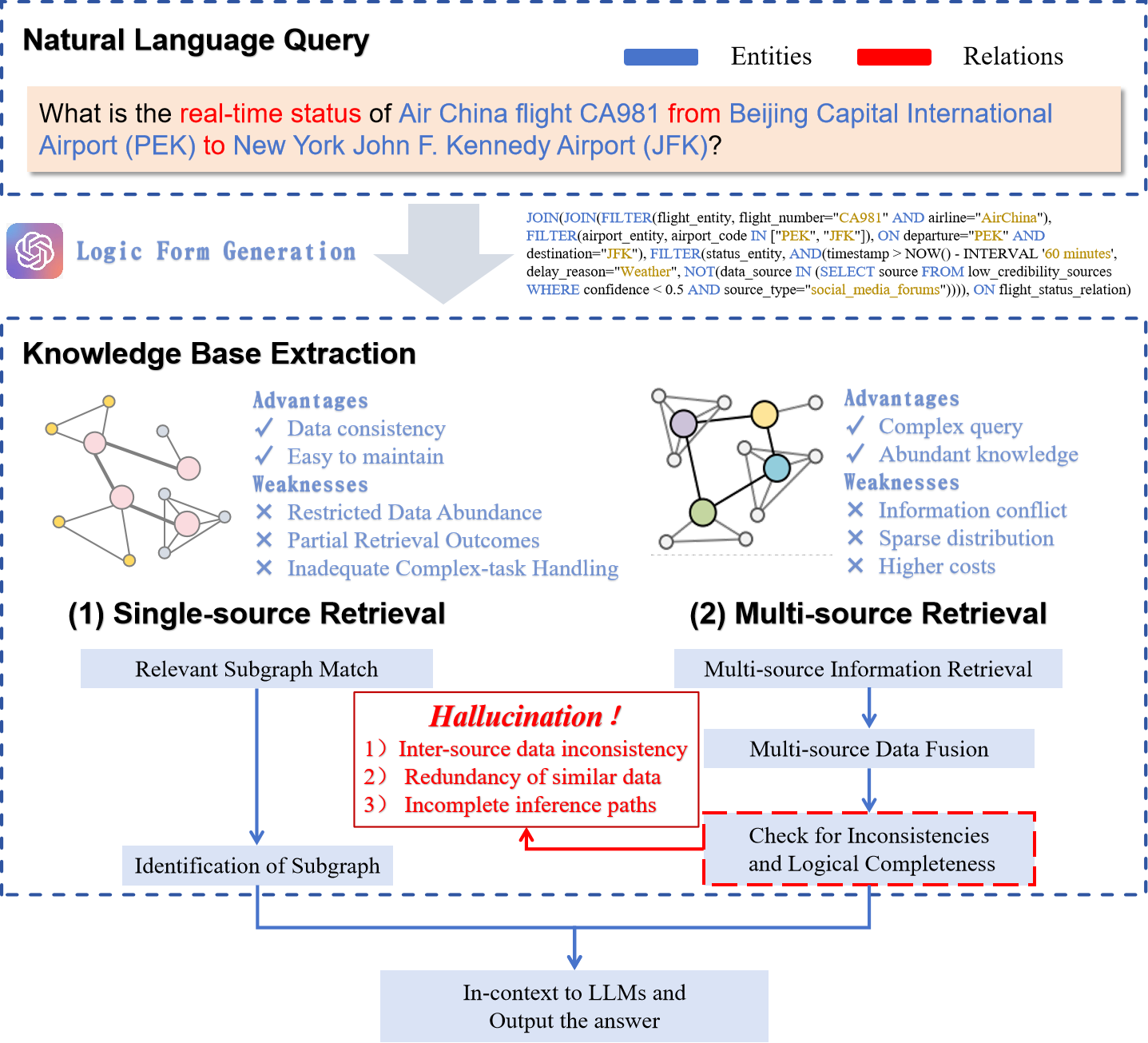}
    \caption{\textcolor{black}{Single-source Retrieval \& Multi-source Retrieval}}
    \label{1}
\end{figure}
% 图1形象地展示了单源和多源数据检索的差异性。其中数据的稀疏分布和不一致性均是多源数据检索特有的问题，会导致LLMs出现严重的幻觉偏差。这也是本文的研究重点。鉴于此，本文专注于缓解多源数据检索下的检索幻觉问题，以赋能知识增强生成。本文主要探讨以下两个挑战：
% (1) 多源数据分布稀疏: 在多元数据检索场景中，查询往往需要提取多个数据源中的检索片段。由于数据存储的格式差异性以及稀疏性，知识元素之间的连接性低，使得rag系统无法有效捕捉跨源数据的逻辑关联，从而影响检索结果的召回率和质量。

% (2) 源间数据不一致：另一方面，由于知识表达的多样性，多源数据的检索片段内容往往不一致的，可能导致检索过程中出现信息冲突，进而影响响应结果的准确性。特别是在领域特定的复杂推理和多跳问答任务中，这种影响更为显著。
Fig. \ref{1} vividly illustrates the differences between single-source and multi-source data retrieval \textcolor{black}{through CA981 flight analysis}. The sparse distribution and inconsistency of data are unique issues in multi-source data retrieval, leading to severe hallucination bias in LLMs. Against this backdrop, we focus on addressing the issue of retrieval hallucinations in multi-source data retrieval to empower knowledge-augmented generation. \textcolor{black}{This work} primarily explores the following two fundamental challenges:
\begin{enumerate}
\item \textbf{Sparse Distribution of Multi-source Data}: \textcolor{black}{Multi-domain queries require fusing structured (SQL tables), semi-structured (JSON logs), and unstructured data (text reports).} Due to the variability in data storage formats and sparsity, the connectivity between knowledge elements is low, making it difficult for RAG systems to effectively capture logical associations across sources, thereby affecting the recall rate and quality of retrieval results.

\item \textbf{Inter-source Data Inconsistency}: \textcolor{black}{Conversely, the inherent diversity in knowledge representations across multi-source data often leads to inconsistencies in retrieved fragments.} These discrepancies may induce information conflicts during retrieval processes, thereby compromising response accuracy. This challenge becomes particularly pronounced in domain-specific complex reasoning and multi-hop question answering tasks.
\end{enumerate}

% 为了应对这些挑战，我们提出了一种创新的方法——MultiRAG，旨在缓解多源知识增强生成中的幻觉现象。首先，我们引入多源线图对异构知识源的快速聚合，以解决数据稀疏分布带来的问题；进而，基于这些整合后的多源线图，我们提出多层级置信度计算方法，以缓解源间数据不一致带来的幻觉问题。这种方法不仅提高了查询效率，还增强了结果的准确性，为多源知识增强生成领域提供了一种有效的解决方案。

% (1) 多源知识聚合：在知识构建模块中，我们引入多源线图作为数据结构对多个查询相关的数据源进行快速聚合和知识结构重建，有效捕捉Chunk文本中的源间数据依赖关系，以此来提供一个统一且集中的多源知识表示。

% (2) 多层级置信度计算：在检索模块中，我们对提取出的知识子图进行图级别和节点级别的置信度计算，目的是筛选并剔除低质量子图和不一致的检索节点，最终提高嵌入上下文的文本质量，以缓解检索幻觉。

% (3) 实验验证与性能比较：我们在现有的多源异构检索数据集和两个复杂Q&A数据集上进行广泛的实验，并与现有的SOTA方法进行比较，证明了所提方法在检索性能上的鲁棒性和准确性。特别是在多源数据检索任务中，我们的方法显著优于其它方法，平均检索的F1分数提高21.6\%，最好性能下的幻觉消除率达xx\%。

To address these issues above, we propose MultiRAG, a novel framework designed to mitigate hallucination in multi-source retrieval augmented generation through knowledge-guided approaches. Initially, we introduce multi-source line graphs for rapid aggregation of knowledge sources to tackle issues arising from sparse data distribution. Subsequently, based on these integrated multi-source line graphs, we propose a multi-level confidence calculation method to ensure the reliability of multi-source data queries. This approach not only enhances query efficiency but also strengthens the accuracy of results, providing an effective solution for the \textcolor{black}{multi-source knowledge-guided RAG}.

The contributions of this paper are summarized as follows:
\begin{enumerate}
%keywords
    \item \textbf{Multi-source Knowledge Aggregation}: In the knowledge construction module, we introduce multi-source line graphs as a data structure for rapid aggregation and reconstruction of knowledge structures from multiple query-relevant data sources. This effectively captures inter-source data dependencies within chunk texts, thereby providing a unified and centralized representation of multi-source knowledge.
    \item \textbf{Multi-level Confidence Calculation:} In the retrieval module, we perform graph-level and node-level confidence calculations on the extracted knowledge subgraphs. The aim is to filter out and eliminate low-quality subgraphs and inconsist retrieval nodes, ultimately enhancing the quality of text embedded in context to alleviate retrieval hallucinations.
    \item \textbf{Experimental Validation and Performance Comparison}: We conducted extensive experiments on existing multi-source retrieval datasets and two complex Q\&A datasets, comparing our approach with existing state-of-the-art(SOTA) methods. This demonstrated the robustness and accuracy of our proposed method in retrieval performance. Particularly in multi-source data retrieval tasks, our method significantly outperforms other SOTA methods by more than 10\%.
\end{enumerate}

\section{PRELIMINARY}
% 在知识引导的检索增强生成（RAG）领域，核心挑战是如何构建高效的检索知识体系，并实现鲁棒的检索性能。本节将从这一角度出发，阐述基本的工作内容，并提供具体的问题形式化定义。
%in this section ...
In the field of Knowledge-Guided RAG, the primary challenges include efficiently accessing relevant knowledge and achieving reliable retrieval performance. \textcolor{black}{This section introduces the core elements of our approach and precisely defines the problems we address.}

% 设 $Q = {q_1, q_2, \ldots, q_n}$ 为查询集合，其中每个$q_i$代表一个查询实例。
% 设 $E = {e_1, e_2, \ldots, e_m}$ 为知识图谱实体集合，其中每个$e_j$代表一个实体。
% 设 $R = {r_1, r_2, \ldots, r_p}$ 为知识图谱关系集合，其中每个$r_k$代表一个关系。
% 设 $D = {d_1, d_2, \ldots, d_t}$ 为文档集合，其中每个$d_l$代表一个文档。我们定义知识引导的检索增强生成问题如下：
Let $Q = \{q_1, q_2, \ldots, q_n\}$ be the set of query instances, where each $q_i$ corresponds to a distinct query. Let $E = \{e_1, e_2, \ldots, e_m\}$ be the set of entities in the knowledge graph, where each $e_j$ represents an entity. Let $R = \{r_1, r_2, \ldots, r_p\}$ be the set of relationships in the knowledge graph, where each $r_k$ represents a relationship. Let $D = \{d_1, d_2, \ldots, d_t\}$ be the set of documents, where each $d_l$ represents a document. We define the knowledge-guided retrieval enhancement generation problem as follows:

\begin{equation}
    \arg\max_{d_i\in D}LLM(q_i, d_i), \sum_{e_j\in E}\sum_{r_k \in R}KG(e_j, r_k, d_i)
\end{equation}

% 其中，$\text{LLM}(q_i, d_l)$ 表示LLM对查询$q_i$和文档$d_l$之间相关性的评分，$\text{KG}(e_j, r_k, d_l)$ 表示实体$e_j$、关系$r_k$与文档$d_l$之间的匹配程度。
\noindent
where $\text{LLM}(q_i, d_l)$ denotes the score of the relevance between query $q_i$ and document $d_l$ assessed by the LLM, and $\text{KG}(e_j, r_k, d_l)$ represents the degree of match between entity $e_j$, relationship $r_k$, and document $d_l$.

\begin{figure}[t]
    \centering
    \includegraphics[width=0.9\linewidth]{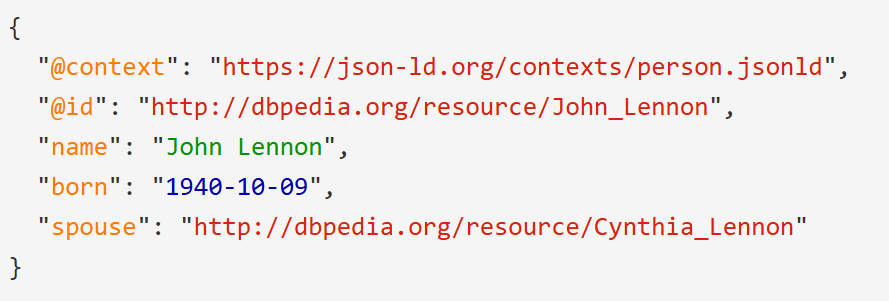}
    \caption{Data format of JSON-LD}
    \label{3}
\end{figure}

\textcolor{black}{Furthermore, we optimize the knowledge construction and retrieval modules by introducing multi-source line graphs to accelerate knowledge establishment and enhance retrieval robustness.} Specifically, the proposed approach is formally defined as follows:
% 进一步地，本文针对KAG过程中的知识构建模块和检索模块进行优化，包括引入了多源线图来加速知识建立过程，并提高检索过程的鲁棒性。具体，本文给出以下定义形式：

% 定义1．跨域数据统一标识．给定在一组异构源H上存在的数据$ D={id,d,name,jsc,meta,(cols\_index)}$．其中$id$表示规范化的唯一标识符，$d$表示该数据文件所在域，$name$表示数据文件名称，$meta$表示文件元数据，$jsc$表示使用JSON-LD存储的文件内容．若所存储数据为结构化数据或其他一些可以使用列式存储模型的数据格式，还会额外存储其所有属性的列索引$cols\_index$，便于快速的检索查询．
\textbf{Definition 1. Multi-source data fusion.} Given a set of sources $H$, the data $D = \{d, name, c, meta\}$ exists, where $d$ represents the domain of data, $c$ represents the content of the data file, $name$ represents the file/attribute name, and $meta$ represents the file metadata. Through a multi-source data fusion algorithm, we can obtain normalized data $\hat{D} = \{id, d, name, jsc, meta, (cols\_index)\}$. Here, $id$ represents the unique identifier for normalization, $d$ indicates the domain where the data file is located, $name$ denotes the data file name, $meta$ denotes the file metadata, and $jsc$ denotes the file content stored using JSON-LD. If the stored data is structured data or other data formats that can use a columnar storage model, the column index $cols\_index$ of all attributes will also be stored for rapid retrieval and query. Fig. \ref{3} provides an example of JSON-LD format.

% 定义2．多源线图[52]．给定一个多源异构知识图谱\mathcal{G}和一个转换后的知识图\mathcal{G'}（Multi-source line graph, MLG），多元线图图满足以下特征：
% \begin{enumerate}
%     \item 
%     % \mathcal{G'}中的一个节点表示中的一个三元组．
%     \item 
%     % \mathcal{G'}中的任意两个节点出现一条关联边当且仅当这两个节点所表示的三元组中存在公共节点．
% \end{enumerate}
% 根据定义可以推断出MLG实现了相关节点的高度聚合，这可以在很大程度上提高数据检索的效率，加速后续的检索和查询算法．
\textbf{Definition 2. Multi-source line graph}\cite{fionda2020learning}. Given a multi-source knowledge graph $\mathcal{G}$ and a transformed knowledge graph $\mathcal{G'}$ (multi-source line graph, MLG), the MLG satisfies the following characteristics:
\begin{enumerate}
    \item A node in $\mathcal{G'}$ represents a \textcolor{black}{triplet}.
    \item There is an associated edge between any two nodes in $\mathcal{G'}$ if and only if the triples represented by these two nodes share a common node.
\end{enumerate}

Based on the definition, it can be inferred that MLG achieves high aggregation of related nodes, which can greatly improve the efficiency of data retrieval and accelerate subsequent retrieval and query algorithms.

% 定义3．跨域同源数据．对于$\mathcal{G}$中的任意两个节点$\upsilon_1$和$\upsilon_2$，定义它们跨域同源当且仅当它们在一次检索中隶属于同一个检索候选集中．
\textbf{Definition 3. Multi-source homologous data}. For any two nodes $\mathcal{\upsilon}_1$ and $\mathcal{\upsilon}_2$ in $\mathcal{G}$, they are defined as multi-source homologous if and only if they belong to the same retrieval candidate set in a single search.

% 定义4．同源节点和同源子图．对于知识图谱\mathcal{G}中的一组跨域同源数据$SV = \{ \upsilon_i \}_{i=1}^n$，定义同源中心节点$snode = \{ name, meta, num, C(v) \}$，同源节点集$U_{snode}$，同源边集$E_{snode}$．其中$name$表示共同属性名，$meta$表示相同的文件元数据，$num$表示同源数据个数，$C(v)$表示数据置信度．定义$snode$与$\upsilon_i$的关联边$e_i=\{ w_i \}_{i=1}^n$，$w_i$表示节点$\upsilon_i$在数据置信度计算中所占权重．则同源中心节点和$S\mathcal{G}$构成同源子图$subS\mathcal{G}$．
\textbf{Definition 4. Homologous node and homologous subgraph}. Given a set of mult-domain homologous data $SV = \{ \upsilon_i \}_{i=1}^n$ in the knowledge graph $\mathcal{G}$, we define the homologous center node as $snode = \{ name, meta, num, C(v) \}$, the set of homologous nodes as $U_{snode}$, and the set of homologous edges as $E_{snode}$. Here, $name$ represents the common attribute name, $meta$ denotes the identical file metadata, $num$ indicates the number of homologous data instances, and $C(v)$ represents the data confidence. We define the association edge between $snode$ and $\upsilon_i$ as $e_i = \{ w_i \}_{i=1}^n$, where $w_i$ represents the weight of node $\upsilon_i$ in the data confidence calculation. Thus, the homologous center node and $S\mathcal{G}$ together form the homologous subgraph $subS\mathcal{G}$.

\begin{figure*}
  \centering
  \includegraphics[width=0.9\textwidth]{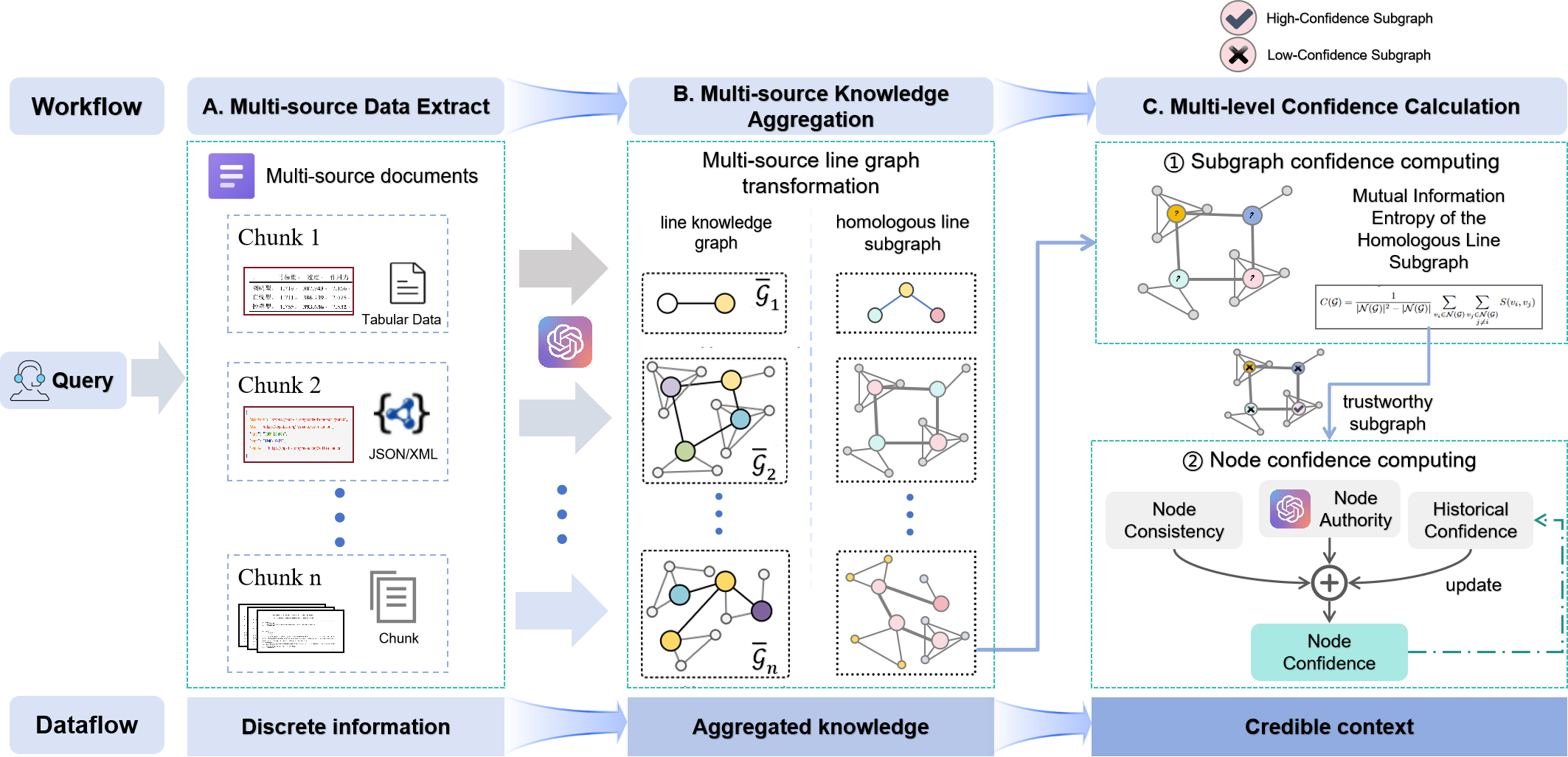}
  \caption{Framework of MultiRAG, including three modules.}
  \label{2}
\end{figure*}

% 定义5．同源线图．对于知识图谱$\mathcal{G}$中的所有同源子图，构成同源知识图谱$S\mathcal{G}$．将同源知识图谱进行线性图转换，得到同源线图$S\mathcal{G'}$．
\textbf{Definition 5. Homologous triple line graph}. For all homologous subgraphs within the knowledge graph $\mathcal{G}$, they collectively constitute the homologous knowledge graph $S\mathcal{G}$. By performing a linear graph transformation on the homologous knowledge graph, we obtain the homologous triple line graph $S\mathcal{G'}$.

% 通过构造同源线图，本文将跨域同源数据聚合到同一个子图中，以同源节点为核心，实现同源数据的快速一致性检验与冲突反馈．此外，知识图谱中也存在大量孤立节点（即不存在同源数据），这些节点也会被加入到同源线图中．
By constructing a homologous triple line graph, multi-source homologous data are aggregated into a single subgraph, centered around homologous nodes, enabling rapid consistency checks and conflict feedback for homologous data. Additionally, the knowledge graph contains a significant number of isolated nodes (i.e., nodes without homologous data), which are also incorporated into the homologous triple line graph.

% 定义6．候选图置信度和候选节点置信度．对于知识图谱$\mathcal{G}$的一次查询$Q(q,\mathcal{G})$，获取对应的同源线图$S\mathcal{G'}$. 候选图置信度是对候选同源子图进行置信度的估算，以评估候选图整体的可信程度；候选节点置信度则是对单个节点进行置信度的评估，以确定其单个属性节点的可信程度。

\textbf{Definition 6. Candidate graph confidence and candidate node confidence.} For a query $Q(q, \mathcal{G})$ on the knowledge graph $\mathcal{G}$, the corresponding Homologous line graph $S\mathcal{G'}$ is obtained. The candidate graph confidence is an estimation of the confidence in the candidate Homologous subgraph, assessing the overall credibility of the candidate graph; the candidate node confidence is an assessment of the confidence in individual node to determine the credibility of single attribute node.

\section{Methodology}
\subsection{Framework of MultiRAG}

% 该章节阐述 MultiRAG的实现思路. 如图2所示，第一步对多源异构数据进行分块和抽取，并构造相应的MLG，实现多源数据初步聚合；第二步需要对MLG进行重构和子图抽取，提取候选同源子图，保证同源数据的一致性存储，便于后续的幻觉评估；第三步，对候选子图的图级别置信度和节点级别置信度分别进行计算，剔除低质量节点以提高回答的可信度，并将抽取到的可信子图返回LLM中形成最终回答．最后，本文还将上述步骤进行整合，形成多源线图提示算法MKLGP．
This section elaborates on the implementation approach of MultiRAG. As shown in Fig. \ref{2}, the first step involves segmenting and extracting multi-source data to construct the corresponding MLG, achieving preliminary aggregation of multi-source data; the second step requires reconstructing the MLG and performing subgraph extraction to identify candidate homologous subgraphs, ensuring consistent storage of homologous data for subsequent hallucination assessment; the third step involves calculating the graph-level and node-level confidence of the candidate subgraphs, eliminating low-quality nodes to enhance the credibility of the response, and returning the extracted trustworthy subgraphs to the LLM to form the final answer. \textcolor{black}{Finally, integrating the aforementioned steps to form the Multi-source Line Graph Prompting algorithm, MKLGP.}

\subsection{Multi-source Line Graph Construction}
% MultiRAG方法首先采用适配器结构来融合跨域异质数据，并统一其存储格式。针对实际应用场景，本文直接从非同源的多种数据格式中获取数据，并将其转化为统一的规范化表示。具体来说，本文收集数据的文件名、文件元数据，并划分文件所在的域。随后，本文将数据内容进行解析，并使用JSON-LD格式进行存储，转化为链接数据。最后，本文赋予数据唯一标识符，形成规范化的数据。
The MultiRAG method initially employs an adapter structure to integrate multi-source data and standardize its storage format. For practical application scenarios, \textcolor{black}{data is directly obtained from various non-homologous formats and transformed into a unified, normalized representation. Specifically, file names and metadata are parsed, and the domains to which the files belong are categorized. Subsequently, the data content is parsed and stored in JSON-LD format, thereby transforming it into linked data.} Finally, unique identifiers are assigned to the data, resulting in normalized datasets. 

% 具体来说，文章为每一种不同的数据格式都设计适配器来进行数据解析．适配器的实现框架大致相同，但需要区分结构化数据、半结构化数据和非结构化数据的解析．对于结构化数据，其解析相当于使用JSON格式存储表格数据，文件中的属性变量采用列式存储模型（decomposition storage model, DSM）进行存储，以便利用列索引提取所有属性信息以进行一致性检验．对于半结构化数据，其解析相当于使用JSON格式存树形数据，文件中采用多层嵌套JSON格式进行存储．该格式数据不存在列式索引和快速检索，需要采用树/图检索算法进行检索(如DFS)．最后，对于非结构化数据，本文目前仅考虑文本数据，直接进行内容存储，并在后续考虑采用LLM进行实体/关系抽取任务来获得相应的信息．
\textcolor{black}{Specifically, a unique adapter is designed for each distinct data format to facilitate data parsing.} Although the implementation frameworks of these adapters are largely similar, it is essential to differentiate between the parsing processes for structured, semi-structured, and unstructured data.

For structured data, parsing involves storing tabular information in JSON format, where attribute variables within the file are managed using a Decomposition Storage Model (DSM). This approach enables the extraction of all attribute information for consistency checks through the use of column indices. In the case of semi-structured data, parsing corresponds to storing tree-shaped data in JSON format with multi-layer nested structures. This data format lacks column indices and does not support fast retrieval, necessitating the use of tree or graph retrieval algorithms, such as DFS, for efficient searching. Finally, for unstructured data, the focus is currently limited to textual information, which is stored directly. Subsequent steps involve leveraging LLMs for entity and relationship extraction tasks to obtain the relevant information.

% 最终，多源异质数据的融合可以表达为以下公式：
The final integration of multi-source data can be expressed by the following formula:
\begin{equation}
\begin{aligned}
    D_{Fusion} = \bigcup_{i=1}^{n} A_i(D_i)
\end{aligned}
\end{equation}

% 其中，$A_i \in \{Ada_{stru}, Ada_{semi-s}, Ada_{unstru}\}$，分别表示结构化数据、半结构化数据和非结构化数据的适配器解析函数．$D_i \in \{D_{stru}, D_{semi-s}, D_{unstru}\}$分别表示结构化数据、半结构化数据和非结构化数据的原始数据集．

% 通过适配器解析后的数据$D_{Fusion} = \{E_{Fu}, R_{Fu}\}$，本文进一步抽取关键信息，并将其链接到知识图谱中。这一过程涉及到实体识别、关系抽取和属性提取等多个步骤，主要采用开源框架OpenSPG来完成。以下公式描述了数据抽取过程：
\noindent
where $A_i \in \{Ada_{\text{stru}}, Ada_{\text{semi-s}}, Ada_{\text{unstru}}\}$, representing the adapter parsing functions for structured data, semi-structured data, and unstructured data, respectively. $D_i \in \{D_{\text{stru}}, D_{\text{semi-s}}, D_{\text{unstru}}\}$ represents the original datasets of structured data, semi-structured data, and unstructured data, respectively.

Through the parsed data $D_{\text{Fusion}} = \{E_{\text{q}}, R_{\text{q}}\}$, \textcolor{black}{we} further extracts key information and links it to the knowledge graph. 
The knowledge construction process involves three key phases implemented through the \textcolor{black}{OpenSPG framework\footnote{\url{https://github.com/OpenSPG/openspg}}\cite{liang2024kag, KGFabric}, in which we use the \texttt{Custom Prompt} module\footnote{\url{https://openspg.yuque.com/}} to integrate LLM-based knowledge extraction.}

\textcolor{black}{For entity recognition, we utilize the \textbf{ner.py} prompts within the \textbf{kag/builder/prompt/default} directory. We first define relevant entity types in the schema. Then, by adjusting the \textit{example.input} and \textit{example.output} in the \textbf{ner.py} prompts, we guide the LLM-based \texttt{SchemaFreeExtractor} to identify entities accurately.}

\textcolor{black}{In relationship extraction, the \textbf{triple.py} prompts play a crucial role. We define relationships in the schema and use the \textbf{triple\_prompt} in the \texttt{SchemaFreeExtractor}. The \textit{instruction} in \textbf{triple.py} ensures that the extracted Subject-Predicate-Object(SPO) triples are related to the entities in the \textit{entity\_list}, enabling effective relationship extraction.}

\textcolor{black}{Regarding attribute extraction, we rely on the entity standardization prompts in \textbf{std.py}. After entity recognition, the \textit{std\_prompt} in the \texttt{SchemaFreeExtractor} standardizes the entities and helps in extracting their attributes. We modify the \textbf{example.input}, \textit{example.named\_entities}, and \textit{example.output} in \textbf{std.py} according to our data characteristics to optimize the attribute extraction process. Through these steps of customizing and applying OpenSPG's prompts, we achieve efficient knowledge extraction.}

The following formula describes the data extraction process:
\begin{equation}
    KB = \sum_{D_i}(\{e_1, e_2,...,e_m\}\bigsqcup\{r_1, r_2,...,r_n\})
\end{equation}
% 其中，$KB$表示初步抽取的知识集合，以数据源作为一级索引，$e_i$表示识别的实体，$r_i$表示抽取的关系。

\subsection{Homologous Subgraph Matching}
% 同源子图匹配．完成信息的初步抽取后，下一步是识别跨域同源数据组集合$\mathcal{SV}s$以及孤立点集$\mathcal{LV}s$。该过程从初始化未访问节点集合$\mathcal{U}{\text{unvisited}} = \mathcal{V}$开始，同时设置同源数据组$\mathcal{SV}s = \emptyset$和孤立点集$\mathcal{LV}s = \emptyset$。通过遍历所有节点，检索各个域中的节点信息，对于匹配到的同源数据，构造同源节点$sg_i$及其对应的关联边$e_i$，并分别加入到同源节点集$\mathcal{U}{sg}$和边集$\mathcal{E}{sg}$中。遍历结束后，将$(\mathcal{U}{sg}, \mathcal{E}{sg})$加入到$\mathcal{SV}s$中。若遍历一轮后未获得同源数据，则将该节点加入孤立点集$\mathcal{LV}s$。遍历完成后，该节点将从$\mathcal{U}{\text{unvisited}}$集合中移除。同源子图匹配的时间复杂度为$O(n\log{n})$，其中$n$为知识图谱$\mathcal{G}$中的节点数量。
After the preliminary extraction of information, the next step is to identify the multi-source homologous data group set $\mathcal{SV}s$ and the isolated point set $\mathcal{LV}s$. This process begins by initializing the unvisited node set $\mathcal{U}_{\text{unvisited}} = \mathcal{V}$, while setting the homologous data group $\mathcal{SV}s = \emptyset$ and the isolated point set $\mathcal{LV}s = \emptyset$. By traversing all nodes and retrieving node information from various domains, for matched homologous data, construct the homologous node $sg_i$ and its corresponding associated edge $e_i$, and add them to the homologous node set $\mathcal{U}_{sg}$ and edge set $\mathcal{E}_{sg}$, respectively. After the traversal, add $(\mathcal{U}_{sg}, \mathcal{E}_{sg})$ to $\mathcal{SV}s$. If no homologous data is obtained after one round of traversal, add the node to the isolated point set $\mathcal{LV}s$. After the traversal is completed, the node will be removed from the $\mathcal{U}_{\text{unvisited}}$ set. The time complexity of homologous subgraph matching is $O(n\log{n})$, where $n$ is the number of nodes in the knowledge graph $\mathcal{G}$. 

% 对于$\mathcal{SV}s$中的每一个同源子图，本文利用同源节点集$\mathcal{U}{sg}$和同源边集$\mathcal{E}{sg}$构造同源线性知识子图$sub\mathcal{S}\mathcal{G'}_i$。随后，将所有的$sub\mathcal{S}\mathcal{G'}_i$和孤立点集$\mathcal{LV}s$进行聚合，得到同源线性知识图$S\mathcal{G'}$。需要注意的是，$S\mathcal{G'}$仅用于同源数据的一致性检验和检索查询，对于其他类型的查询，本文仍然在原始知识图谱$\mathcal{G}$中进行。
\textcolor{black}{For each homologous subgraph in $\mathcal{SV}s$, homologous linear knowledge subgraph $sub\mathcal{S}\mathcal{G'}_i$ is constructed by utilizing the homologous node set $\mathcal{U}_{sg}$ and the homologous edge set $\mathcal{E}_{sg}$.} Subsequently, all $sub\mathcal{S}\mathcal{G'}_i$ and the isolated point set $\mathcal{LV}s$ are aggregated to obtain the homologous linear knowledge graph $S\mathcal{G'}$. It should be noted that $S\mathcal{G'}$ is solely used for consistency checks and retrieval queries of homologous data; \textcolor{black}{other types of queries still conducts operations on the original knowledge graph $\mathcal{G}$.}

\begin{figure}[t]
    \centering
    \includegraphics[width=0.9\linewidth]{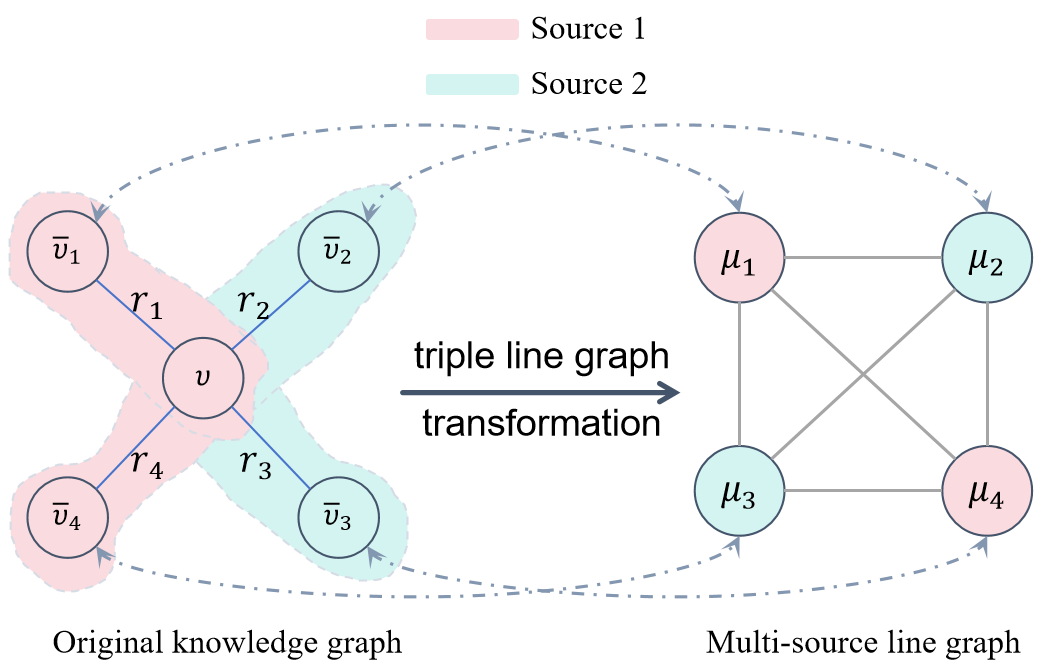}
    \caption{Example of multi-source line graph transformation}
    \label{4}
\end{figure}

% 这里，本文给出一个同源线图的简单样例．如图4所示，同源节点与4个同源数据相关联，转化为线性知识图后，形成一个四阶完全图，表明四个三元组之间两两同源．
Here, we provide a simple example of a homologous triple line graph. As shown in Fig. \ref{4}, a homologous node is associated with 4 homologous data points. After transformation into a triple line graph, it forms a complete graph of order 4, indicating that the four triples are pairwise homologous.

\subsection{Multi-level Confidence Computing}
% 本文将一次检索中来自于不同域的候选数据定义为跨域同源数据．这些数据已经被抽取到同源线图中进行临时的存储。虽然它们面向同一查询对象，但往往会给出不一致的参考答案。考虑到每次查询可能发生的检索错误不尽相同，本文采用了两阶段的置信度计算方法，首先计算单个同源线图的置信度，再计算每个候选节点的置信度，来确定最终答案的候选集。
\textcolor{black}{We define the candidate data} from different domains obtained in a single retrieval as multi-source homologous data. These data have been extracted into a homologous line graph for temporary storage. Although targeting the same query object, they often provide inconsistent reference answers. Considering the varying retrieval errors, the multi-level confidence calculation method is adpoted \textcolor{black}{in this framework}. First, the confidence of individual homologous line graphs is calculated, followed by the confidence of each candidate node, to determine the final set of answer candidates.

\subsubsection{Graph-Level Confidence Computing}

% 阶段一引入了一种基于互信息熵的置信度计算方法，用以评估同源线图的置信度。这种方法的核心思想是，如果同源线图中的两个同属性节点在内容上接近一致，则它们的相似度高，从而置信度也高；反之，则置信度低。
In the first stage, a confidence calculation method based on mutual information entropy is introduced to assess the confidence of homologous line graphs. The core idea of this method is that if two nodes with the same attributes in a homologous line graph are close in content, their similarity is high, and thus their confidence is also high; conversely, if they are not, their confidence is low.

% 设$\mathcal{G}$为同源线图，$\mathcal{N}(\mathcal{G})$为图中的节点集合。对于任意两个同属性的节点$v_i, v_j \in \mathcal{N}(\mathcal{G})$，定义它们之间的相似度$S(v_i, v_j)$基于互信息熵的计算方法。互信息熵$I(v_i, v_j)$衡量了两个节点属性内容的相互依赖性，其计算公式为：
Let $\mathcal{G}$ be a homologous line graph, and $\mathcal{N}(\mathcal{G})$ be the set of nodes in the graph. For any two nodes $v_i, v_j \in \mathcal{N}(\mathcal{G})$ with the same attributes, the similarity $S(v_i, v_j)$ between them is defined based on the calculation method of mutual information entropy. The mutual information entropy $I(v_i, v_j)$ measures the interdependence of the attribute content of the two nodes, and its calculation formula is:
\begin{equation}
    I(v_i,v_j) = \sum_{x\in V_i}\sum_{y\in V_j}p(x,y)\log(\frac{p(x,y)}{p(x)p(y)})
\end{equation}
% 其中，$V_i$和$V_j$分别是节点$v_i$和$v_j$的属性值集合，$p(x, y)$是$v_i$取属性值$x$和$v_j$取属性值$y$的联合概率分布，$p(x)$和$p(y)$分别是$x$和$y$的边缘概率分布。
where $V_i$ and $V_j$ are the sets of attribute values for nodes $v_i$ and $v_j$, respectively, $p(x, y)$ is the joint probability distribution of $v_i$ taking attribute value $x$ and $v_j$ taking attribute value $y$, and $p(x)$ and $p(y)$ are the marginal probability distributions of $x$ and $y$, respectively.

The similarity $S(v_i, v_j)$ can be defined as the normalized form of mutual information entropy to ensure that its value lies within the interval [0, 1]:

% 相似度$S(v_i, v_j)$可以定义为互信息熵的归一化形式，以确保其值在[0, 1]区间内：
\begin{equation}\label{s1}
    S(v_i,v_j) = \frac{I(v_i,v_j)}{H(V_i)+H(V_j)}
\end{equation}
% 其中，$H(V_i)$和$H(V_j)$分别是节点$v_i$和$v_j$属性值集合的熵，计算公式为：
where $H(V_i)$ and $H(V_j)$ are the entropies of the attribute value sets of nodes $v_i$ and $v_j$, respectively, calculated as:
\begin{equation}
    H(V) = -\sum_{x\in V}p(x)\log p(x)
\end{equation}

% 进而，同源线图$\mathcal{G}$的置信度$C(\mathcal{G})$可以通过计算图中所有节点对的相似度$S(v_i, v_j)$的平均值来确定：
Subsequently, the confidence $C(\mathcal{G})$ of the homologous line graph $\mathcal{G}$ can be determined by calculating the average similarity $S(v_i, v_j)$ of all node pairs in the graph:
\begin{equation}
    C(\mathcal{G}) = \frac{1}{|\mathcal{N}(\mathcal{G})|^2 - |\mathcal{N}(\mathcal{G})|}\sum_{v_i\in \mathcal{N}(\mathcal{G})}\sum_{\substack{v_j\in \mathcal{N}(\mathcal{G})\\j\neq i}}S(v_i,v_j)
\end{equation}
% 其中，$|\mathcal{N}(\mathcal{G})|$表示图中节点的数量。高置信度的同源线图表明其内部节点在属性内容上具有高度一致性。
where $|\mathcal{N}(\mathcal{G})|$ denotes the number of nodes in the graph. \textcolor{black}{ Notably, a homologous line graph exhibiting high confidence demonstrates that its constituent nodes maintain strong attribute-level consistency across their content representations.}

\subsubsection{Node-Level Confidence Computing}

% 第二阶段将计算单个节点的置信度，该计算综合考虑了节点的一致性、权威性和历史置信度。以下是详细的计算方法和公式。
In the second phase, \textcolor{black}{the confidence of individual node $C(v)$} is calculated, which takes into account the node's consistency, authority, and historical confidence. The following are the detailed calculation methods and formulas.

\paragraph{Node Consistency Score} The node consistency score $S(v)$ reflects the consistency of the node across different data sources. We use mutual information entropy to calculate the similarity between node pairs, thereby assessing consistency. For a node $v$, its consistency score can be expressed as:
% 节点一致性得分。节点一致性得分$S(v)$反映了节点在不同数据源中的一致性。我们使用互信息熵来计算节点对之间的相似度，从而评估一致性。对于节点$v$，其一致性得分可以表示为：
\begin{equation}
    S_{n}(v) = \frac{1}{|N(v)|}\sum_{u\in N(v)}S(v,u)
\end{equation}
% 其中，$N(v)$是与节点$v$同属性的节点集合，$S(v,u)$是公式\ref{s1}。
where $N(v)$ is the set of nodes with the same attributes as node $v$, and $S(v, u)$ is the similarity between nodes $v$ and $u$ as defined in Equation \ref{s1}.

\begin{algorithm}[t]
\caption{Multi-level Confidence Computing Algorithm}
\label{HKGM}
\begin{algorithmic}[1]
\Procedure{Confidence\_Computing}{$v, D$}
    \State $S_n(v) \leftarrow$ \textit{Equation} (8)
    \State $AuthLLM(v) \leftarrow $\textit{Equation} (10)
    \State $Authhist(v) \leftarrow $\textit{Equation} (11)
    \State $A(v) \leftarrow$\textit{Equation} (9)
    \State $C(v) \leftarrow S_n(v) + A(v)$
    \State \Return $C(v)$
\EndProcedure

\Procedure{MCC}{$\mathcal{G}, Q, D$}
    \State $\mathcal{SV}s \leftarrow \emptyset, \mathcal{LV}s \leftarrow \emptyset$ 
    \State $\mathcal{U}_{unvisited} \leftarrow V$ 
    \While{$\mathcal{U}_{unvisited} \neq \emptyset$}
        \State $v \leftarrow$ pop a node from $\mathcal{U}_{unvisited}$
        \ForAll{$D \in D$}
            \If{$v \in Data(Q, subSG_i)$}
                \State $C(v) \leftarrow$ Confidence\_Computing$(v, D)$
                \If{$C(v) > \theta$}
                    \State $\mathcal{U}_{sg} \leftarrow \mathcal{U}_{sg} \cup \{v\}$
                    \State $\mathcal{E}_{sg} \leftarrow \mathcal{E}_{sg} \cup \{e_i\}$
                \Else
                    \State $\mathcal{LV}s \leftarrow \mathcal{LV}s \cup \{v\}$
                \EndIf
            \EndIf
        \EndFor
        \If{$\mathcal{U}_{sg} \neq \emptyset$}
            \State $\mathcal{SV}s \leftarrow \mathcal{SV}s \cup (\mathcal{U}_{sg}, \mathcal{E}_{sg})$
            \State $\mathcal{U}_{sg} \leftarrow \emptyset, \mathcal{E}_{sg} \leftarrow \emptyset$
        \EndIf
    \EndWhile
    \State \Return $\mathcal{SV}s, \mathcal{LV}s$
\EndProcedure

\end{algorithmic}
\end{algorithm}

% 节点权威性得分。本文将权威性得分划分为LLM评估的节点权威性和节点的历史权威性两个部分。该得分反应节点的重要性以及真实性。本文通过专家大模型（LLM）来综合评估节点的权威性。节点的权威性得分$A(v)$可以通过以下公式计算：
\paragraph{Node Authority Score} \textcolor{black}{Authority score is divided into two parts}: the node's authority assessed by the LLM and the node's historical authority. This score reflects the importance and authenticity of the node. \textcolor{black}{Additionally, we use an expert LLM to comprehensively evaluate the authority of the node}. The node's authority score $A(v)$ can be calculated using the following formula:
\begin{equation}\label{eq9}
    A(v) = \alpha\cdot Auth_{LLM}(v) + (1-\alpha)\cdot Auth_{hist}(v)
\end{equation}
where $\alpha$ is a weight coefficient that balances the contributions of LLM-assessed authority and historical authority, satisfying $0 \leq \alpha \leq 1$.
% 其中，$\alpha$是权重系数，用于平衡LLM评估的权威性和历史权威性的贡献，满足$0 \leq \alpha \leq 1$。

% 受益于PTCA模型[引用]对知识可信度的计算思想，本文基于节点的全局影响力和局部连接强度进行$Auth_{\text{LLM}}(v)$评估。LLM可以综合实体之间的关联强度、实体类型信息以及多步路径信息对知识的可信度进行计算。
Benefiting from the calculation idea of knowledge credibility in the PTCA\cite{ptca}, \textcolor{black}{$Auth_{\text{LLM}}(v)$ is assessed by the global influence and local connection strength of the node}. The LLM can comprehensively calculate the credibility of knowledge by integrating the association strength between entities, entity type information, and multi-step path information.

\begin{equation}
    Autm_{LLM}(v) = \frac{1}{1+e^{-\beta\cdot C_{LLM}(v)}}
\end{equation}
where $C_{\text{LLM}}(v)$ is the authority score provided by the LLM for node $v$ is the average value of all nodes' $C_{\text{LLM}}(v)$, and $\beta$ is a parameter that controls the steepness of the scoring curve.
% 其中，$C_{\text{LLM}}(v)$是LLM为节点$v$提供的权威性评分，$\mu$是所有节点$C_{\text{LLM}}(v)$的平均值，$\beta$是控制评分曲线陡峭程度的参数。

\paragraph{Historical Authority} $Auth_{\text{hist}}(v)$ is an authority score based on the node's historical data. \textcolor{black}{Inspired by Zhu's work \cite{zhu2024fusionquery}, we expect to} use the credibility of historical data sources and current query-related data for incremental estimation.
% 历史权威性$Auth_{\text{hist}}(v)$是基于节点历史数据的权威性评分，受到先前工作\cite{zhu2024fusionquery}的启发，本文期望利用历史数据源的可信度和当前查询相关的数据进行增量估计．

\begin{equation}
    Auth_{hist}(v) = \frac{\mathcal{H}\cdot Pr^h(D)+\sum_{\upsilon_p \in D_\upsilon [q]}Pr(\upsilon_p)}{\mathcal{H} + |Data(q, subS\mathcal{G'}_i)|}
\end{equation}
where $\mathcal{H}$ is the number of entities provided by data source $D$ for all historical queries, $Pr^h(D)$ is the historical credibility of data source $D$, $D_\upsilon [q]$ is the set of correct answers, and $Data(q, subS\mathcal{G'}_i)$ is the query-related data obtained from the multi-source line subgraph.
% 其中，$\mathcal{H}$是数据源$D$为所有历史查询提供的实体数，$Pr^h(D)$是数据源$D$的历史可信度，$D_\upsilon [q]$是正确答案集，$Data(q, subS\mathcal{G'}_i)$是从源的线性知识子图中获取的查询相关数据。

\begin{algorithm}[t]
\caption{Multi-source Knowledge Line Graph Prompting}
\label{MKLGP}
\begin{algorithmic}[1]
\Procedure{MKLGP}{$q$}
    \State $E_q, R_q \leftarrow \text{Logic Form Generation}(q)$
    \State $D_q \leftarrow \text{Multi Document Extraction}(V_q)$
    \State $\mathcal{SG'} \leftarrow \text{Prompt}(D_q)$
    \State $\mathcal{SV}s, \mathcal{LV}s \leftarrow \text{MCC}(\mathcal{SG'}, q, D_q)$
    \State $C_{\text{nodes}}, \mathcal{G}_{A} \leftarrow \text{Prompt}(\mathcal{SV}s, \mathcal{LV}s)$ 
    \State $Answer \leftarrow \text{Generating Trustworthy Answers}(C_{\text{nodes}}, \mathcal{G}_{A})$ 
    \State \Return $Answer$
\EndProcedure
\end{algorithmic}
\end{algorithm}
% 最终，本文设计可信同源子图匹配算法MCC，对同源子图的数据源可信度进行计算，以保证嵌入LLM的知识图质量．算法2的实现流程如下．需要注意的是，MCC算法并没有直接给出最终的图置信度和节点置信度，还需要经过Prompt\ref{prompt2}才能得到最终的结果。

Ultimately, we designed the multi-level confidence computing algorithm, MCC, to calculate the credibility of the data sources in the homologous subgraph, ensuring the quality of the knowledge graph embedded in the LLM. The algorithm is shown in Algorithm\ref{HKGM}. 

It should be noted that the MCC algorithm does not directly provide the final graph confidence and node confidence; these values must be obtained through \textcolor{black}{prompt} to achieve the ultimate results.

\subsection{Multi-source knowledge line graph prompting}
% 本文提出 Multi-source Knowledge Line Graph Prompting (MKLGP) algorithm for multi-source data retrieval. 给定一个用户的查询$q$，本文首先使用LLM对$q$进行意图抽取以及实体和关系的抽取，并生成对应的逻辑关系。接下来，对数据集进行多文档的初步筛查，获得文本片段（片段用Chunk，括号里不要翻译），并构造多源线图进行知识聚合。进一步，匹配同源子图，并使用MCC算法获得可信的查询节点集和孤立点集 $\mathcal{SV}s, \mathcal{LV}s$。最后，通过提示词 \ref{Prompt2}来获得图置信度，进一步计算节点置信度来增强答案的可信度，并最终嵌入到LLM的上下文中，生成可信的检索答案。
\textcolor{black}{We propose the Multi-source Knowledge Line Graph Prompting (MKLGP) algorithm for multi-source data retrieval. Given a user query $q$, LLM is firstly employed to extract the intent, entities, and relationships from $q$, and generates the corresponding logical relationships. The dataset then undergoes multi-document filtering to derive text chunks, followed by constructing a Multi-source Line Graph (MLG) for knowledge aggregation.} Further, it matches homogeneous subgraphs and utilizes the MCC algorithm to obtain a set of credible query nodes and isolated points $\mathcal{SV}s, \mathcal{LV}s$. Finally, by leveraging the \textcolor{black}{prompt}, the graph confidence is obtained, and the node confidence is calculated to enhance the credibility of the answer. The results are then embedded into the context of the LLM to generate a credible retrieval answer.

\section{Experiments}
% 本章节将对多源知识融合以及多层级置信度计算这两个模块进行实验和性能分析。文章将与其它先进的RAG方法、数据融合问答方法和KBQA方法的基线进行比较，通过广泛的实验来评估本文方法所使用的技术的鲁棒性和高效性．
This section will conduct experiments and performance analysis on the construction of homologous line graphs and the multi-level confidence calculation modules. \textcolor{black}{Baseline methods will be compared with other SOTA multi-document retrieval QA methods, data fusion methods, and KBQA methods. Extensive experiments will be conducted to assess the robustness and efficiency of MultiRAG,} which aims to answer the following questions.

% 相较于其它的基础和SOTA的数据检索模型，MultiRAG的检索召回性能如何？(在数据集上的正常实验效果)
% 加入多层级置信度计算后，MultiRAG在多跳问答数据集中的表现如何？(hop=1, 2, 3 使用GPT-4还是Reranking model 作为评估打分器)
% 数据稀疏性(不用多源线图)和数据不一致性(幻觉增多)对检索召回质量的影响？
% MultiRAG对各类幻觉的缓解效果如何？
% MultiRAG(多源知识融合和多层级置信度计算)的时间开销如何？
% Case Study

% How does the retrieval recall performance of MultiRAG compare with other foundational and SOTA data retrieval models?
% What are the respective impacts of data sparsity and data inconsistency on the quality of retrieval recall?
% How is the performance of MultiRAG in multi-hop Q\&A datasets after incorporating multi-level confidence calculation?
% How effective are the two modules of MultiRAG individually?
% What are the time costs of the various modules in MultiRAG?

% 待补充实验：LLM/Reranking model as expert evaluator，LLM as LLM as evaluator prompts.

\begin{itemize}
    \item \textbf{Q1:} How does the retrieval recall performance of MultiRAG compare with other data fusion models and SOTA data retrieval models?
    \item \textbf{Q2:} What are the respective impacts of data sparsity and data inconsistency on the quality of retrieval recall?
    \item \textbf{Q3:} How effective are the two modules of MultiRAG individually?
    \item \textbf{Q4:} How is the performance of MultiRAG in multi-hop Q\&A datasets after incorporating multi-level confidence calculation?
    \item \textbf{Q5:} What are the time costs of the various modules in MultiRAG?
\end{itemize}

\subsection{Experimental Settings}
% 数据集．为了验证多源线图构建的高效性和对检索性能的提升，文章在四个真实世界的基准数据集上进行多源数据融合实验\cite{li2015truth,dong2009integrating,yin2007truth}．（1）电影数据集包含从13个来源收集的电影数据．（2）图书数据集包含来自10个来源的图书数据．（3）航班数据集从20个来源收集了1200多个航班的信息．（4）股票数据集收集来自20个来源的1000个股票符号的交易数据．在实验中，我们对四个数据集各发出100个查询，来验证其检索的高效性．

% 值得注意的是，Movies数据集和Flights数据集相对稠密，而Books数据集和Stocks数据集相对稀疏，这会对模型效果造成影响。

% 210，100，260，100
\paragraph{\textbf{Datasets}} To validate the efficiency of multi-source line graph construction and its enhancement of retrieval performance, the article conducts multi-source data fusion experiments on four real-world benchmark datasets \cite{li2015truth,dong2009integrating,yin2007truth}, as is shown in Table \ref{dataset}. (1) The movie dataset comprises movie data collected from 13 sources. (2) The book dataset includes book data from 10 sources. (3) The flight dataset gathers information on over 1200 flights from 20 sources. (4) The stock dataset collects transaction data for 1000 stock symbols from 20 sources. In the experiments, we issue 100 queries for each of the four datasets to verify their retrieval efficiency.

It is noteworthy that the Movies dataset and the Flights dataset are relatively dense, while the Books dataset and the Stocks dataset are relatively sparse, which can impact the model's performance.

% 另外的，为了验证MultiRAG在复杂Q&A数据集上的检索鲁棒性，我们选择了两个多跳问题回答数据集HotpotQA和2WikiMultiHopQA。这两个数据集都是基于维基百科文档构建的，允许我们使用一致的文档语料库和检索器为LLMs提供外部引用。考虑到实验成本的限制，我们从每个实验数据集的验证集中对300个问题进行了子样本分析,这些问题都至少包含了。
Additionally, to validate the robustness of the MultiRAG on complex Q\&A datasets, we selected two multi-hop question answering datasets, HotpotQA\cite{yang2018hotpotqa} and 2WikiMultiHopQA\cite{2wiki}. Both datasets are constructed based on Wikipedia documents, allowing us to utilize a consistent document corpus and retriever to provide external references for LLMs. Considering the constraints of experimental costs, we conducted a subsample analysis on 300 questions from the validation sets of each experimental dataset.

\begin{table}[H]
\caption{Statistics of the datasets preprocessed}
\label{dataset}
\centering
\scriptsize % 设置表格字体大小为小
\renewcommand{\arraystretch}{1.5} % 设置行间距为默认的1.2倍
\begin{tabular}{cccccc}
\hline
\textbf{Datasets} & \textbf{Data source} & \textbf{Sources} & \textbf{Entities} & \textbf{Relations} & \textbf{Queries} \\
\hline
\multirow{3}{*}{Movies} & JSON(J) & 4 & 19701 & 45790 & \multirow{3}{*}{\color{black}100} \\
                        & KG(K) & 5 & 100229 & 264709 & \\
                        & CSV(C) & 4 & 70276 & 184657 & \\
\hline
\multirow{3}{*}{Books} & JSON(J) & 3 & 3392 & 2824 & \multirow{3}{*}{100} \\
                      & CSV(C) & 3 & 2547 & 1812 & \\
                      & XML(X) & 4 & 2054 & 1509 & \\
\hline
\multirow{2}{*}{Flights} & CSV(C) & 10 & 48672 & 100835 & \multirow{2}{*}{\color{black}100} \\
                        & JSON(J) & 10 & 41939 & 89339 & \\
\hline
\multirow{2}{*}{Stocks} & CSV(C) & 10 & 7799 & 11169 & \multirow{2}{*}{100} \\
                       & JSON(J) & 10 & 7759 & 10619 & \\
\hline
\end{tabular}
\end{table}

\paragraph{\textbf{Evaluation Metrics}}
% 评价指标．为了评估多源线图融合的有效性，我们参考先前的实验指标[58,61]，使用F1分数作为数据融合结果的评价指标，即精度（precise, P）和召回率（recall, R）的调和平均值，计算公式为．
To assess effectiveness, we adopt the F1 score as the evaluation metric for the data fusion results, following previous experimental metrics \cite{lin2018domain,wang2024knowledge,yin2007truth, zhao2012bayesian}. The F1 score is the harmonic mean of precision (P) and recall (R), calculated as follows:
\begin{equation}
     F1 = 2 \times \frac{P \times R}{P + R} 
\end{equation}

% 同时，为了评估MKLGP算法的检索可信度，我们采用召回率Recall@K，对子图筛选前、节点筛选前以及节点筛选后的三个阶段进行计算。此外，我们还使用查询响应时间T（以秒为单位）作为一个评价指标来验证知识聚合的效率．
Furthermore, to evaluate the retrieval credibility of MKLGP Algorithm, we utilize the recall metric, specifically Recall@K, to assess performance at three distinct stages: before subgraph filtering, before node filtering, and after node filtering. In addition, we employ the query response time \( T \) (measured in seconds) as an evaluative metric to verify the efficiency of knowledge aggregation.

\begin{table*}[t]
\caption{Comparison with baseline methods and SOTA methods for multi-source knowledge fusion}
\label{TQ15}
\renewcommand{\arraystretch}{1.5}
\centering{
\scalebox{0.95}{
\begin{tabular}{c|c|cccc|cccccccc|cc}
\hline
\multirow{3}{*}{Datasets} & \multirow{3}{*}{\begin{tabular}[c]{@{}c@{}}Data\\ source\end{tabular}} & \multicolumn{4}{c|}{Data Fusion Methods (Baseline)}  & \multicolumn{8}{c|}{SOTA Methods} & \multicolumn{2}{c}{Our Method}  
\\ \cline{3-16}
& & \multicolumn{2}{c}{TF} & \multicolumn{2}{c|}{LTM} & \multicolumn{2}{c}{IR-CoT} & \multicolumn{2}{c}{MDQA} & \multicolumn{2}{c}{ChatKBQA}  & \multicolumn{2}{c}{FusionQuery} & \multicolumn{2}{|c}{MCC}  
\\ \cline{3-16}
& & F1/\% & Time/s & F1/\% & Time/s & F1/\% & Time/s & F1/\% & Time/s & F1/\% & Time/s & F1/\% & Time/s & F1/\% & Time/s 
\\ \hline
\multirow{4}{*}{Movies}   & J/K & 37.1 & 9717 & 41.4 & 1995 & 43.2       & 1567 & 46.2 & 1588 & 45.1 & 3809 & \textbf{53.2}     & 122.4    & \underline{52.6} & \textbf{98.3} 
\\
& J/C & 41.9 & 7214  & 42.9 & 1884 & 45.0 & 1399 & 44.5 & 1360 & 42.7 & 3246  & 52.7 & 183.1 & \textbf{54.3} & \textbf{75.1} 
\\
& K/C & 37.8 & 2199 & 41.2 & 1576 & 37.6 & 1014  & 45.2  & 987 & 40.4 & 2027  & 42.5 & 141.0 & \textbf{49.1} & \textbf{86.0} 
\\
& J/K/C & 36.6 & 11225 & 40.8 & 2346 & 41.5 & 2551 & 49.8 & 2264 & 44.7 & 5151 & 53.6 & \textbf{137.8} & \textbf{54.8} & \underline{157}  
\\ \hline

\multirow{4}{*}{Books} & J/C & 40.2 & 1017 & 42.4 & 195.3s & 35.2       & 147.6 & 55.7 & 124.2  & 56.1 & 165.0  & 58.5 & 22.7 & \textbf{63.5} & \textbf{13.66} 
\\
& J/X & 35.5 & 1070 & 35.6 & 277.7 & 36.1 & 178.7 & 55.1 & 115.6  & 54.7 & 200.1 & 57.9 & 20.6 & \textbf{63.1} & \textbf{13.78}
\\
& C/X & 43.0 & 1033 & 44.1 & 232.6 & 42.6  & 184.5      & 57.2 & 115.6  & 55.6 & 201.4 & \underline{60.3} & 21.5    & \textbf{64.2}   & \textbf{13.54} 
\\
& J/C/X & 37.3 & 2304 & 41.0 & 413.2 & 40.4 & 342.6 & 56.4 & 222.6  & 57.1 & 394.1 & 59.1 & 47.0 & \textbf{66.8} & \textbf{27.4}  
\\ \hline

Flights & C/J & 27.3 & 6049 & 79.1 & 14786 & 58.3 & 214.0  & \textbf{76.5}   & 360  & 76.8 & 376 & 74.2 & \textbf{20.2} & \underline{74.9}  & \underline{80}           
\\ \hline

Stocks & C/J & 68.4 & 2.30 & 19.2 & 1337 & 64.8 & 53.3  & 65.2 & 78.4  & 64.0 & 88.9 & 68.0   & \textbf{0.33} & \textbf{78.6} & \underline{12.1} \\
\hline
\end{tabular}
}}
\begin{tablenotes}
        \footnotesize
        \item \textcolor{black}{* The F1 score is for Q1 and time is for Q5.}
        \item \textcolor{black}{* Bold represents the optimal metrics, while underlined text indicates the sub-optimal metrics. The same applies to the following text.}
\end{tablenotes}
\end{table*}

\paragraph{\textbf{Hyper-parameter Settings}}
% 对于所有的基线，我们根据MultiRAG的特性仔细调整参数．所有的方法都是在Python 3.10和CUDA 11.6环境中实现。除去使用GPT-3.5-Turbo进行CoT的实验，其余工作均使用Llama3-8B 作为基座模型。对于每一类不同的数据格式，我们将其切分为Chunk后，均采用JSON-LD的格式来存储其切片编号和所在数据源，以及变换后的多源线图中的三元组节点，并以此来实现简单的互索引．对于超参数设置，温度参数$\beta$设置为0.5．历史查询的实体数量初始化为50，初始的节点置信度阈值定义为0.7，图置信度阈值为0.5。所有的实验都是在一台拥有Intel(R) Core(TM) Ultra 9 185H 2.30GHz和内存512GB的设备上进行的．
For all baselines, we carefully adjusted the parameters according to the characteristics of MultiRAG. All methods were implemented in a Python 3.10 and CUDA 11.6 environment. Except for the experiments using GPT-3.5-Turbo for CoT, the rest of the work utilized Llama3-8B-Instruct as the base model. For each different data format, after slicing into Chunks, we stored the slice numbers, data source locations, and transformed triple nodes in the multi-source line graph using JSON-LD format, thereby enabling simple cross-indexing. 

For hyperparameter settings, the temperature parameter $\beta$ was set to 0.5. The number of entities in historical queries was initialized to 50, the initial node confidence threshold was defined as 0.7, and the graph confidence threshold was set to 0.5. All experiments were conducted on a device equipped with an Intel(R) Core(TM) Ultra 9 185H 2.30GHz and 512GB of memory.

\paragraph{\textbf{Baseline Models}}
% 基线模型比较。为了证明MultiRAG方法的优越性，本文将它与当前最先进的多文档问答方法（MDQA）和基于知识库的问答方法（KBQA）进行比较。这些方法包括基础方法Chain of Thought (CoT)和标准检索增强生成（Standard RAG），以及以下最新的方法：IRCoT、HybridRAG、MetaRAG（WWW 24'）、ChatKBQA（ACL 24'）、MDQA（AAAI 24'）、GraphRAG、StructRAG和KAG。

% 此外，我们与以下基线方法进行比较：朴素的数据融合方法MajorityVoter（MV）、经典的基于迭代的数据融合方法TruthFinder（TF）[60]、概率数据融合方法LTM [61]。
To demonstrate the superiority of the MultiRAG method, \textcolor{black}{we compare} it with basic data fusion methods and SOTA methods, including the multi-document question-answering methods and knowledge base question-answering methods. 

Thanks to Zhu's work\footnote{https://github.com/JunHao-Zhu/FusionQuery} \cite{zhu2024fusionquery}, we compare with the following baseline methods:
\begin{enumerate}
    \item \textbf{TruthFinder(TF)}\cite{yin2007truth}: the classic iterative data fusion method.
    \item \textbf{LTM}\cite{zhao2012bayesian}: the probabilistic data fusion method.
    \item \textbf{CoT}\cite{cot} is a foundational approach that involves step-by-step reasoning to reach a conclusion, we use GPT-3.5-Turbo as the base model.
    \item \textbf{Standard RAG}\cite{rag1} is a method that combines the strengths of retrieval and generation models to answer questions.
\end{enumerate}

Moreover, we also summerize these SOTA methods below:
\begin{itemize}
        \item \textbf{IRCoT}\cite{ircot} is an advanced method that refines the reasoning process through iterative retrieval.
        \item \textbf{ChatKBQA}\cite{luo2023chatkbqa} is a conversational interface-based method for knowledge base question answering.
        \item \textbf{MDQA}\cite{mdqa} is a method designed to extract answers from multiple documents effectively.
        \item \textbf{FusionQuery}\cite{zhu2024fusionquery} is a SOTA method based on the efficient on-demand fusion query framework.
        \item \textbf{RQ-RAG}\cite{rqrag} is a method that integrates external documents and optimizes the query process to handle complex queries.
        \item \textbf{MetaRAG}\cite{metarag} is a method that employs metacognitive strategies to enhance the retrieval process.
\end{itemize}

\paragraph{\textbf{Dataset Preprocessing}}
% 数据集预处理．为了使数据集更加贴合实际应用场景，同时证明所提出方法对异质数据的适用性，我们将四个数据集进行拆分和重构，划分为表格数据（结构化数据），嵌套JSON数据（半结构化数据）以及XML数据（半结构化数据）三类数据格式，分别存储在csv、json和xml文件格式中．同时也保留了部分直接以KG格式进行存储的数据．表1显示了数据集划分后的详细统计数据．
To better align the datasets with real-world application scenarios and to demonstrate the applicability of the proposed method to multi-source data, we have split and reconstructed the four datasets into three categories of data formats: tabular data (structured data), nested JSON data (semi-structured data), and XML data (semi-structured data), stored respectively in csv, json, and xml file formats. We also retained some data directly stored in KG format. Table \ref{dataset} displays the detailed statistics after the dataset division.

\begin{table*}[t]
\caption{Ablation experiments of multi-source knowledge aggregation(MKA) and multi-level confidence computing(MCC)}
\label{t3}
\renewcommand{\arraystretch}{1.5}
\centering
\begin{tabular*}{\textwidth}{@{\extracolsep{\fill}}c|c|ccc|ccc|ccc|ccc|ccc}
\hline
\multirow{2}{0.045\textwidth}{Datasets} & \multirow{2}{0.045\textwidth}{Source} & \multicolumn{3}{c|}{MultiRAG} & \multicolumn{3}{c|}{w/o MKA} & \multicolumn{3}{c|}{w/o Graph Level} & \multicolumn{3}{c|}{w/o Node Level} & \multicolumn{3}{c}{w/o MCC} \\ 
\cline{3-17} & & \multicolumn{1}{c}{F1/\%} & \multicolumn{1}{c}{QT/s} & \multicolumn{1}{c|}{PT/s} & \multicolumn{1}{c}{F1/\%} & \multicolumn{1}{c}{QT/s} & \multicolumn{1}{c|}{PT/s} & \multicolumn{1}{c}{F1/\%} & \multicolumn{1}{c}{QT/s} & \multicolumn{1}{c|}{PT/s} & \multicolumn{1}{c}{F1/\%} & \multicolumn{1}{c}{QT/s} & \multicolumn{1}{c|}{PT/s} & \multicolumn{1}{c}{F1/\%} & \multicolumn{1}{c}{QT/s} & \multicolumn{1}{c}{PT/s}  \\ \hline
\multirow{4}{*}{Movies}   
&J/K  & 52.6    & 25.7   & 62.64 & 48.2   & 2783   & 62.64 & \textcolor{black}{45.3} & \textcolor{black}{50.1} & \textcolor{black}{58.2} & \textcolor{black}{38.7} & \textcolor{black}{21.3} & \textcolor{black}{0.31} & 31.6 & 25.7 & 0.28 
\\
& J/C & 54.3 & 12.7   & 61.36    & 49.1   & 1882   & 61.36  & \textcolor{black}{46.8} & \textcolor{black}{28.9} & \textcolor{black}{57.4} & \textcolor{black}{40.2} & \textcolor{black}{10.5} & \textcolor{black}{0.29} & 30.5  & 12.7 & 0.29
\\
&K/C & 49.1 & 31.6 & 64.40 & 45.5 & 4233   & 64.40  & \textcolor{black}{42.7} & \textcolor{black}{65.3} & \textcolor{black}{61.8} & \textcolor{black}{35.9} & \textcolor{black}{28.4} & \textcolor{black}{-0.27} & 33.1  & 31.6 & -0.29
\\
& J/K/C & 54.8 & 39.2 & 60.8 & 47.5   & 4437   & 60.8  & \textcolor{black}{48.1} & \textcolor{black}{75.6} & \textcolor{black}{56.2} & \textcolor{black}{41.5} & \textcolor{black}{35.8} & \textcolor{black}{0.30} & 34.7  & 39.2 & 0.32
\\ \hline

\multirow{4}{*}{Books}    
& J/C & 63.5    & 1.19   & 2.47    & 57.1   & 11.9   & 2.47  & \textcolor{black}{55.2} & \textcolor{black}{4.7} & \textcolor{black}{2.12} & \textcolor{black}{49.8} & \textcolor{black}{0.92} & \textcolor{black}{0.18} & 43.4  & 1.19  & 0.22 
\\
& J/X & 63.1 & 1.22   & 2.56    & 59.3   & 11.7   & 2.62  & \textcolor{black}{54.7} & \textcolor{black}{5.1} & \textcolor{black}{2.24} & \textcolor{black}{48.3} & \textcolor{black}{0.89} & \textcolor{black}{0.19} & 42.6  & 1.22  & 0.22
\\ 
& C/X & 64.2 & 1.16   & 2.38    & 55.3   & 8.39   & 2.38  & \textcolor{black}{53.9} & \textcolor{black}{3.9} & \textcolor{black}{2.05} & \textcolor{black}{47.1} & \textcolor{black}{0.85} & \textcolor{black}{0.16} & 41.0  & 1.16   & 0.17 
\\
& J/C/X & 66.8 & 1.31   & 3.07    & 57.2   & 15.8   & 3.08  & \textcolor{black}{59.4} & \textcolor{black}{6.3} & \textcolor{black}{2.89} & \textcolor{black}{52.7} & \textcolor{black}{1.12} & \textcolor{black}{0.21} & 36.4  & 1.31  & 0.20 
\\ \hline

Flights & C/J & 74.9 & 29.8   & 109.9   & 72.2 & \textbf{NAN}   & 109.9   & \textcolor{black}{68.3} & \textcolor{black}{142.7} & \textcolor{black}{98.5} & \textcolor{black}{61.4} & \textcolor{black}{25.3} & \textcolor{black}{0.85} & 52.1 & 29.8 & 1.07
\\ \hline

Stocks & C/J & 78.6 & 2.72   & 5.36    & 69.6   & 450.8  & 5.36  & \textcolor{black}{72.1} & \textcolor{black}{8.9} & \textcolor{black}{4.12} & \textcolor{black}{65.3} & \textcolor{black}{1.98} & \textcolor{black}{0.15} & 45.4 & 2.72 & 0.17  
\\ \hline
\end{tabular*}
\end{table*}

\subsection{Evaluation of Multi-source Knowledge Aggregation (MKA)}

\textbf{Q1: How does the retrieval recall performance of MultiRAG compare with other data fusion models and SOTA data retrieval models?}

% 为了验证MultiRAG中的多源知识聚合模块的有效性，我们采用F1分数和查询时间分别在四个多源异构查询数据集上进行评估。通过不同的基线模型和SOTA模型来替换本文提出的MCC模块，可以获得多组实验结果来评估其多域查询的性能．表2．总结了MKLGP和基线在四个数据集上的数据查询性能，Q1仅关注方法的F1得分，其中包括4种数据融合方法，以及3种支持数据融合的SOTA方法。
To validate the effectiveness of the multi-source knowledge aggregation module (MKA) in MultiRAG, we assess it using F1 scores and query times across four multi-source query datasets. \textcolor{black}{By substituting the fusion query algorithm with different baseline models and SOTA models}, multiple sets of experimental results are botained to evaluate its performance in multi-domain querying. Table \ref{TQ15} summarizes the data querying performance of MKLGP and baselines on the four datasets; Q1 focuses solely on the F1 scores of the methods, which includes four data fusion methods and three SOTA methods that support data fusion.

% 表 \ref{Q15}显示，MCC模块在四个数据集上均优于所有的对比模型．实验结果表明，它比最好的基础数据融合模型的F1分数高了10\%+，并且比其他基线获得了更好的性能．MV方法在所有数据集上的性能都较差，这是因为MV只能为查询返回一个答案，无法适应一个查询通常有多个返回值的情况．例如，一部电影或一本书通常有多个导演或作者．而大部分方法均在Movies数据集和Flights数据集上的表现要显著好于Books数据集和Stocks数据集上。这是因为Movies和Flights数据集本身相对稠密，而之前的SOTA模型会在知识充沛的情况下媲美或优于我们的方法，这是可以接受的。而在较为稀疏的Books和Stocks数据集上，我们的方法相较于SOTA方法平均提升了10\%+。

Table \ref{TQ15} demonstrates that the MCC module outperforms all comparative models across four datasets. Experimental results indicate that it achieves an F1 score that is more than 10\% higher than the best baseline data fusion model and obtains superior performance compared to other baselines. The MV method performs poorly on all datasets because it can only return a single answer for a query, which fails to accommodate the common scenario where a query has multiple return values. For instance, a movie or a book typically has multiple directors or authors. However, the majority of methods show significantly better performance on the Movies and Flights datasets than on the Books and Stocks datasets. This is because the Movies and Flights datasets are inherently denser, and previous SOTA models can match or outperform our approach in situations where knowledge is abundant, which is acceptable. In contrast, on the more sparse Books and Stocks datasets, our method achieves an average improvement of more than 10\% over SOTA methods.

\textbf{Q2: What are the respective impacts of data sparsity and data inconsistency on the quality of retrieval recall?}

% 为了验证MultiRAG是针对文章所提出的两个挑战的高效解决方案，我们从以下两个角度出发进行试验：(1) 多源数据的稀疏性：为了验证MultiRAG在稀疏数据场景下具有更鲁棒的性能，我们将四个预处理过后的数据集，进行30\%、50\%和70\%的随机关系MASK遮盖，使得数据间的联系更加稀疏，但仍然保证查询答案是可被检索的。(2) 多源数据的一致性：为了验证MultiRAG在多源数据场景下对数据一致性具有良好的检验能力，我们将四个预处理过后的数据集，进行进行30\%、50\%和70\%的三元组增量添加(新增的三元组是原有三元组的复制)，并对新增三元组的关系边进行完全的打乱，以破坏多源数据的一致性．之后，我们采用MultiRAG对两种扰动方案下的数据集进行实验。
\textcolor{black}{MultiRAG demonstrates good robustness in scenarios of varying data sparsity and inconsistency. To validate it, we conducted experiments from the following two perspectives}. 1) Sparsity of multi-source data: We applied 30\%, 50\%, and 70\% random relationship masking to four pre-processed datasets, making the connections between data sparser while ensuring that the query answers are still retrievable. 2) Consistency of multi-source data: We added 30\%, 50\%, and 70\% of triple increments (the new triples are copies of the original triples) to the four pre-processed datasets, and completely shuffled the relationship edges of the added triples to disrupt the consistency of multi-source data. Subsequently, we employed MultiRAG to experiment with datasets under both perturbation schemes.

\begin{table}[t]
\centering
\caption{Performance comparison on HotpotQA and 2WikiMultiHopQA datasets}
\renewcommand{\arraystretch}{1.5}
\label{Recall-Halu}
{\scalebox{1}{
 \begin{tabular}{ccccc}
\toprule
{\multirow{2}{*}{\textbf{Method}}} & \multicolumn{2}{c}{{\textbf{HotpotQA}}} & \multicolumn{2}{c}{{\textbf{2WikiMultiHopQA}}} \\
\cmidrule{2-5}
& \textbf{Precision} & \textbf{Recall@5} & \textbf{Precision} & \textbf{Recall@5} \\
\midrule
Standard RAG & 34.1 & 33.5 & 25.6 & 26.2 \\
GPT-3.5-Turbo+CoT & 33.9 & 47.2 & 35.0 & 45.1 \\
IRCoT & 41.6 & 41.2 & 42.3 & 40.9 \\
ChatKBQA & 47.8 & 42.1 & 46.5 & 43.7 \\
MDQA & 48.6 & \underline{52.5} & 44.1 & 45.8 \\
RQ-RAG & \underline{51.6} & 49.3 & 45.3 & 44.6 \\
MetaRAG & 51.1 & 49.9 & \underline{50.7} & \underline{52.2} \\
MultiRAG & \textbf{59.3} & \textbf{62.7} & \textbf{55.7} & \textbf{61.2} \\
\bottomrule
\end{tabular}
}}
\end{table}

\begin{figure*}[t]
    \centering
        \begin{subfigure}[b]{0.24\textwidth}
        \includegraphics[width=\textwidth]{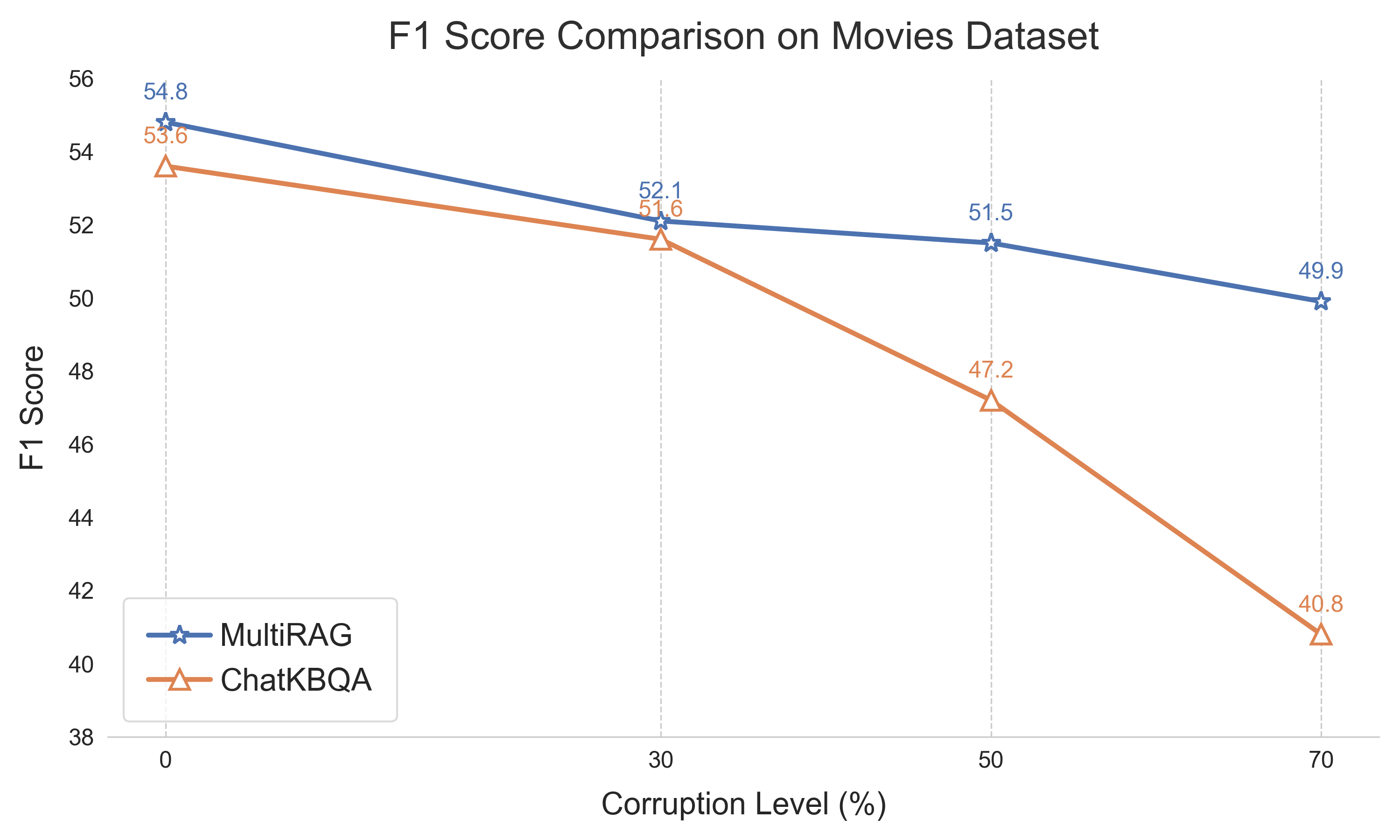}
        \caption{F1 Score in Movies}
        \label{moviesvs}
        \end{subfigure}
    \hfill
    \begin{subfigure}[b]{0.24\textwidth}
        \includegraphics[width=\textwidth]{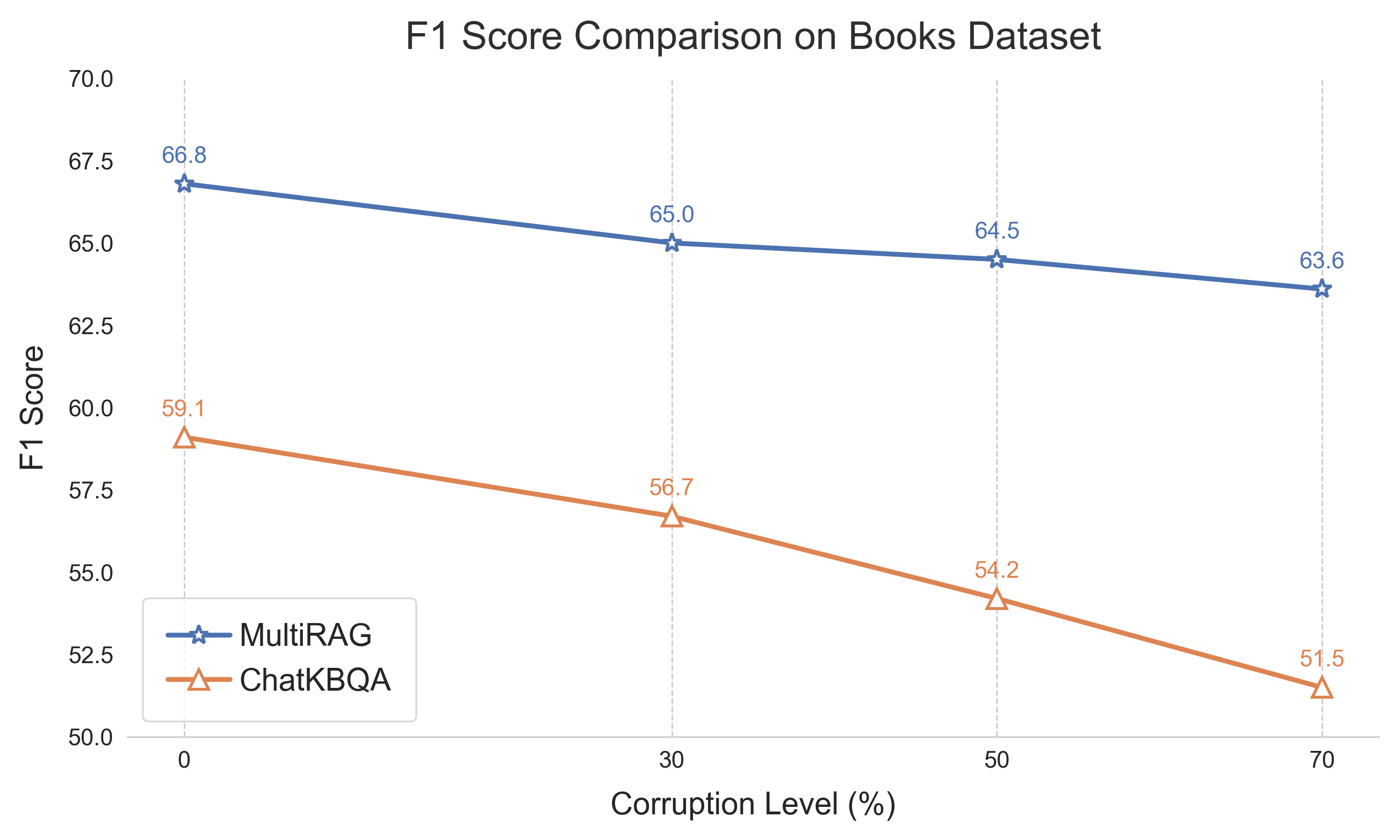}
        \caption{F1 Score in Books}
        \label{bookvs}
    \end{subfigure}
    \hfill
    \begin{subfigure}[b]{0.24\textwidth}
        \includegraphics[width=\textwidth]{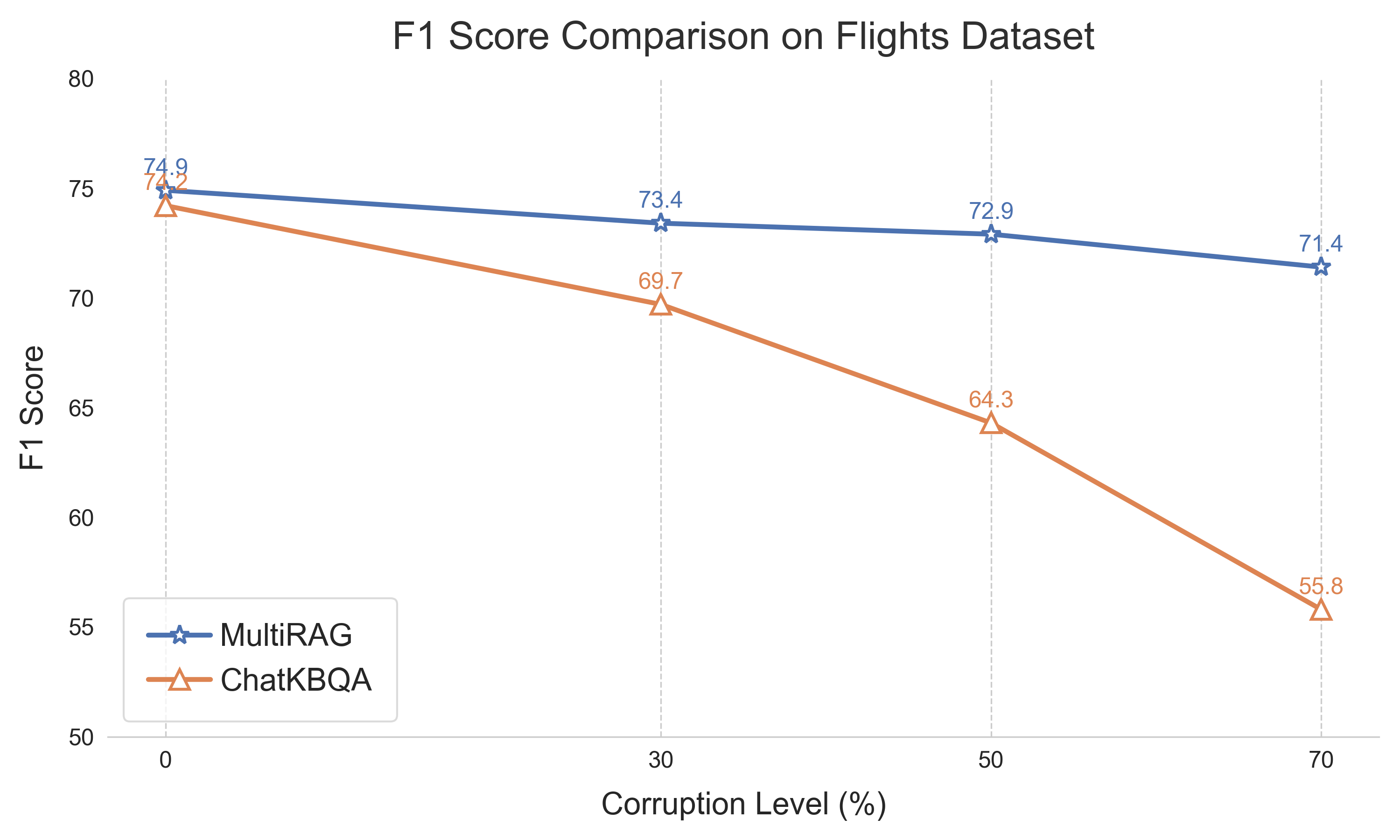}
        \caption{F1 Score in Flights}
        \label{flightvs}
    \end{subfigure}
    \hfill
        \begin{subfigure}[b]{0.24\textwidth}
        \includegraphics[width=\textwidth]{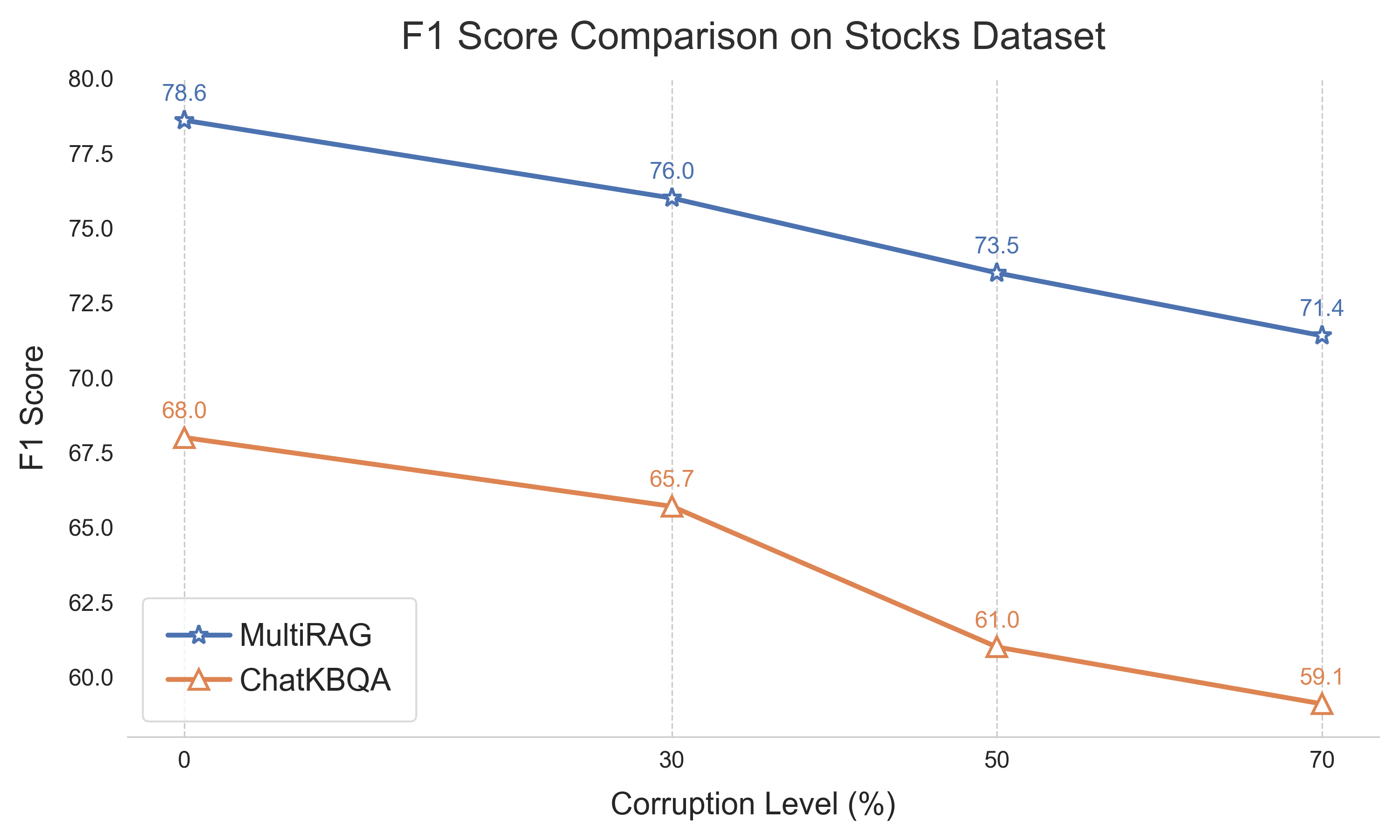}
        \caption{F1 Score in Stocks}
        \label{stockvs}
    \end{subfigure}

    \caption{Experimental results of Q2, where (a) and (b) display the multi-source data sparsity experiments, and (c) and (d) display the multi-source data consistency experiments.}
    \label{Q31}
\end{figure*}

\begin{figure*}[t]
    \centering
    % \begin{subfigure}[b]{0.48\textwidth}
    %     \includegraphics[width=\textwidth]{Figures/Movie_F1_Trend_Analysis.png}
    %     \caption{F1 Scores in Movies Dataset from different sources}
    %     \label{movieF1}
    % \end{subfigure}
    % \hfill
    % \begin{subfigure}[b]{0.48\textwidth}
    %     \includegraphics[width=\textwidth]{Figures/Book_F1_Trend_Analysis.png}
    %     \caption{F1 Scores in Books Dataset from different sources}
    %     \label{bookF1}
    % \end{subfigure}
    \begin{subfigure}[b]{0.48\textwidth}
        \includegraphics[width=\textwidth]{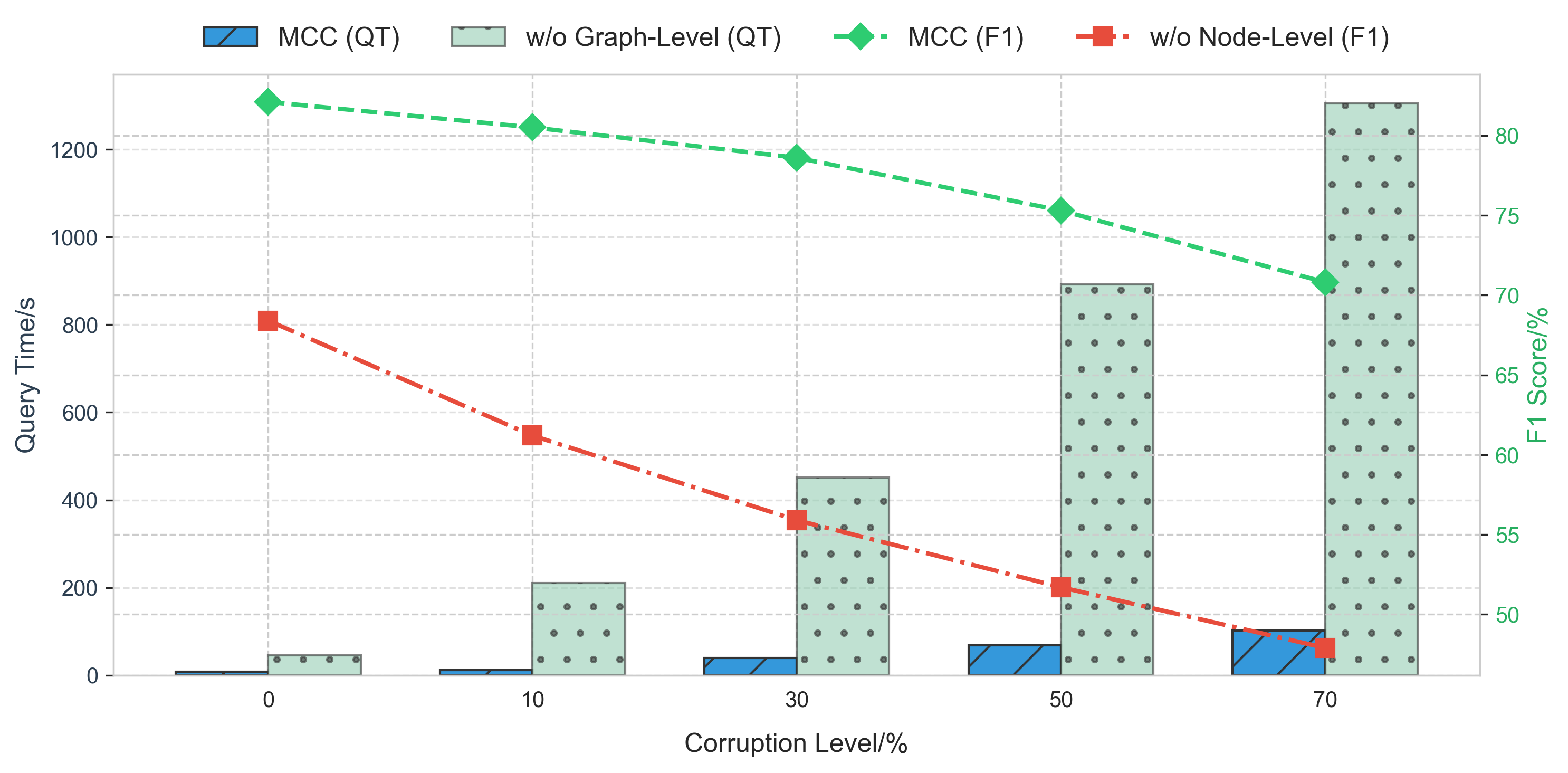}
        \caption{{\color{black}Efficiency-Accuracy Tradeoff of Movies Dataset}}
        \label{movieF1}
    \end{subfigure}
    \hfill
    \begin{subfigure}[b]{0.48\textwidth}
        \includegraphics[width=\textwidth]{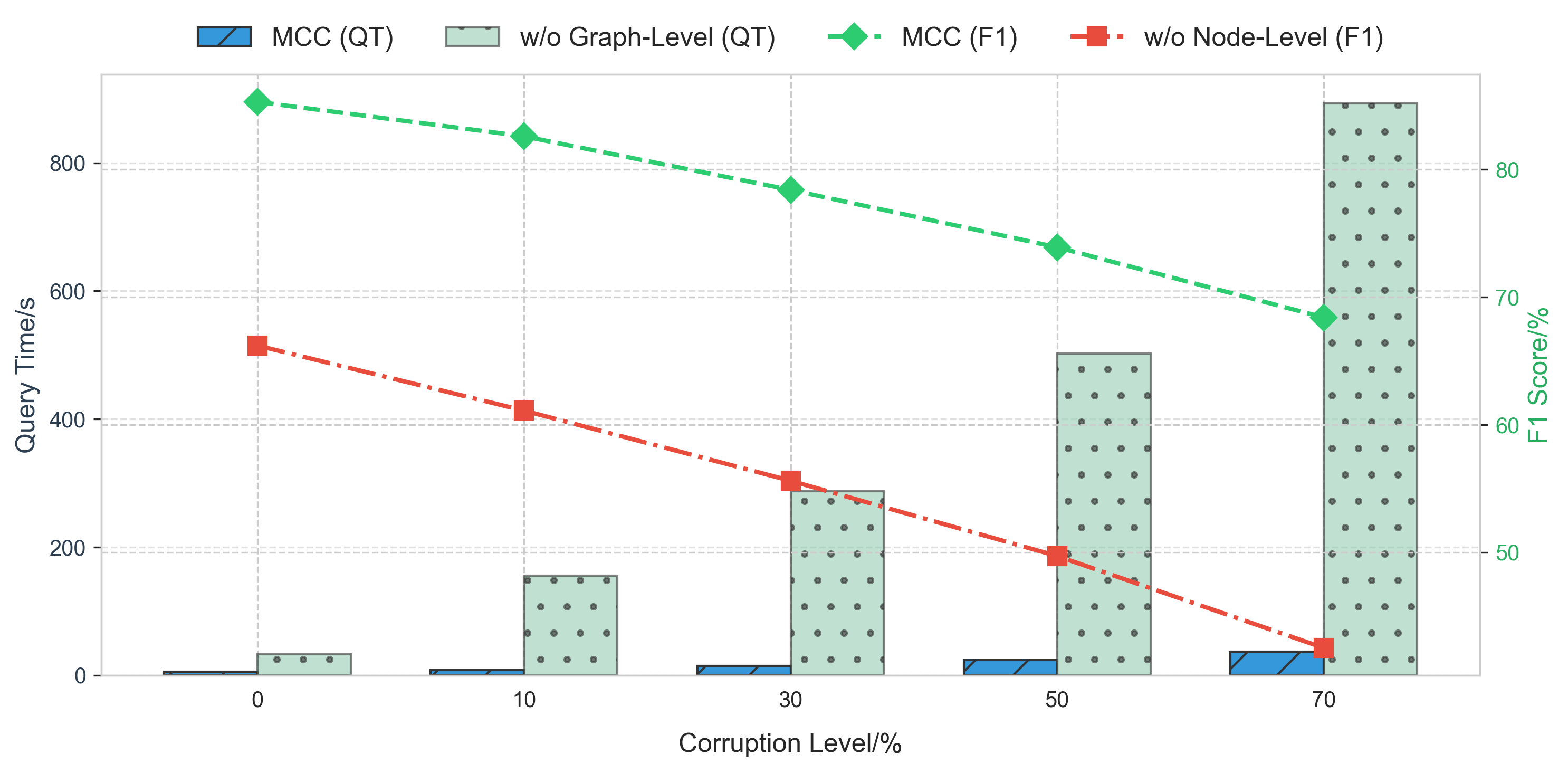}
        \caption{{\color{black}Efficiency-Accuracy Tradeoff of Books Dataset}}
        \label{bookF1}
    \end{subfigure}
    \caption{F1 score and Query Time of Movies and Books with corruption level 0\%, 10\%, 30\%, 50\%, 70\% in different sources}
    \label{Q32}
\end{figure*}
% 首先针对数据稀疏性，我们对MultiRAG（Ours）和ChatKBQA（SOTA）分别开展实验。实验结果表明，MultiRAG在面对数据稀疏性的挑战时，展现出了显著的鲁棒性。具体来说，在经过30%、50%和70%的关系MASK遮盖后，MultiRAG在Books数据集上的F1分数仅从66.8%下降至60.0%，而在Stocks数据集上，其F1分数从78.6%下降至71.0%。这一下降幅度相对较小，说明MultiRAG能够有效地维持其性能，即使在大量关系被遮盖的情况下。

% 相比之下，ChatKBQA在同一条件下的性能下降更为显著。在Books数据集上，ChatKBQA的F1分数从59.1%下降至53.0%，而在Stocks数据集上，其F1分数从68.0%下降至62.0%。这一结果揭示了ChatKBQA在处理稀疏数据时面临的挑战，尤其是在数据连接被大量遮盖的情况下，其性能受到了较大的影响。

% 接下来是多源数据一致性的算法鲁棒性实验。我们对Books和Stocks数据集进行了不同程度的扰动，以测试MultiRAG和ChatKBQA在数据一致性破坏时的性能变化。实验结果显示，MultiRAG在面对数据一致性破坏时展现出了卓越的鲁棒性，而ChatKBQA的性能则在扰动情况下迅速下降。

% 具体来说，在Movies数据集上，我们对原始数据集进行了30%、50%和70%的三元组增量添加，并打乱了新增三元组的关系边。结果显示，MultiRAG的F1分数从54.8\%略微下降至52.1\%、51.5\%和49.9\%，而ChatKBQA的F1分数则从53.6\%显著下降至51.6\%、47.2\%和40.8\%。在Flights数据集上，我们也进行了相同的扰动操作，MultiRAG的F1分数从74.9%小幅下降至73.4%、72.9\%和71.4%，而ChatKBQA的F1分数则从74.2%大幅下降至69.7%、64.3%和55.8%。

% 这些结果表明，即使在数据一致性受到严重破坏的情况下，MultiRAG依然能够保持较高的性能稳定性，而ChatKBQA的性能则对数据一致性的破坏更为敏感。
Firstly, to address data sparsity, we conducted experiments on MultiRAG (Ours) and ChatKBQA (SOTA). The experimental results demonstrate that MultiRAG exhibits significant robustness when faced with the challenge of data sparsity. 

Specifically, after applying 30\%, 50\%, and 70\% relationship masking, the F1 score of MultiRAG on the Books dataset only dropped from 66.8\% to 60.0\%. On the Stocks dataset, its F1 score decreased from 78.6\% to 71.0\%, which have been shown in Fig.\ref{bookvs} and Fig.\ref{stockvs}. These moderate decreases indicate that MultiRAG can effectively maintain its performance even when a substantial number of relationships are masked.

In contrast, ChatKBQA's performance decline under the same conditions is more significant. On the Books dataset, ChatKBQA's F1 score dropped from 59.1\% to 53.0\%, and on the Stocks dataset, its F1 score decreased from 68.0\% to 62.0\%. This outcome reveals the challenges ChatKBQA faces when dealing with sparse data, especially when a large number of data connections are masked, significantly impacting its performance.

Next, we conducted robustness experiments on multi-source data consistency. We perturbed the Books and Stocks datasets to varying degrees to test the performance changes of MultiRAG and ChatKBQA when data consistency is disrupted. The experimental results show that MultiRAG demonstrates excellent robustness in the face of data consistency disruption, while ChatKBQA's performance declines rapidly under perturbation.

Specifically, as is shown in Fig. \ref{moviesvs}, on the Movies dataset, we added 30\%, 50\%, and 70\% triple increments to the original dataset and randomized the relationship edges of the added triples. The results show that MultiRAG's F1 score slightly decreased from 54.8\% to 52.1\%, 51.5\%, and 49.9\%, while ChatKBQA's F1 score significantly dropped from 53.6\% to 51.6\%, 47.2\%, and 40.8\%. On the Flights dataset shown in Fig. \ref{flightvs}, we performed the same perturbation operations, and MultiRAG's F1 score slightly decreased from 74.9\% to 73.4\%, 72.9\%, and 71.4\%, while ChatKBQA's F1 score substantially dropped from 74.2\% to 69.7\%, 64.3\%, and 55.8\%.

These results indicate that even when data consistency is severely compromised, MultiRAG can still maintain a high level of performance stability, whereas ChatKBQA's performance is more sensitive to disruptions in data consistency.

% \begin{figure*}[htbp]
%     \centering
%     \caption{Prompts for Homologous Line Graph Extraction and Multi-level Confidence Computing}
%     \begin{subfigure}[b]{0.9\textwidth}
%     \includegraphics[width=\textwidth]{Figures/prompt1.png}
%     \end{subfigure}
%     \subcaption{Prompts for Homologous Line Graph Extraction}
%     \label{Prompt1}
%     \vspace{0.3cm}
%     \begin{subfigure}[b]{0.9\textwidth}
%     \includegraphics[width=\textwidth]{Figures/prompt2.png}
%     \end{subfigure}
%     \subcaption{Prompts for Multi-level Confidence Computing}
%     \label{Prompt2}
% \end{figure*}

\begin{table*}[t]
\centering
\renewcommand{\arraystretch}{1.5}  % 增加行距，让表格更加饱满
\caption{Case Study}
\label{tab:case_study}
\resizebox{\textwidth}{!}{%
\begin{tabular}{@{\hspace{18pt}} m{4.8cm} p{20.2cm} @{\hspace{8pt}}}
\toprule
\multicolumn{2}{c}{\large\bfseries\textcolor{datagreen}{Query}: "What is the real-time status of Air China flight CA981 from Beijing Capital International Airport (PEK) to New York John F. Kennedy Airport (JFK)?"} \\
\midrule

 \textbf{\large Data Sources} & \\

\textcolor{datagreen}{\large \textbf{Structured}} & \texttt{\large CA981, PEK, JFK, Delayed, 2024-10-01 14:30} \\
\textcolor{datagreen}{\large \textbf{Semi-structured}} & \large \{"flight": "CA981", "delay\_reason": "Weather", "source": "AirChina"\} \\
\textcolor{datagreen}{\large \textbf{Unstructured}} & \large "Typhoon Haikui impacts PEK departures after 14:00." \\
\midrule
\textbf{\large MKA Module} & 
\large \textbf{Structured parsing}: Flight attributes mapping \newline
\large \textbf{LLM extraction}: \texttt{(CA981, DelayReason, Typhoon) @0.87} \\
\midrule
\large \textbf{MLG Subgraph} & 
\begin{center}
\begin{adjustbox}{width=0.95\linewidth, height=3in, keepaspectratio}
\begin{tikzpicture}[
    node distance=1.1cm and 1.8cm,  % 增加节点间距
    every node/.style={rectangle, rounded corners=5pt, draw=black, thick, fill=white, 
                      align=center, font=\sffamily\large, minimum height=1.2cm, minimum width=2.2cm, },
    edge/.style={->, >=Stealth, very thick, black, shorten >=2pt, shorten <=2pt}
]
    \node[fill=moduleblue!20, draw=moduleblue!60, text=black] (CA981) {CA981\\(Flight)};
    \node[below left=of CA981, fill=datagreen!20, draw=datagreen!60] (PEK) {PEK\\(Origin)};
    \node[below right=of CA981, fill=datagreen!20, draw=datagreen!60] (JFK) {JFK\\(Destination)};
    \node[right=of CA981, fill=processorange!20, draw=processorange!60] (Delayed) {Delayed\\(Status)};
    \node[above=of Delayed, fill=red!20, draw=red!60] (OnTime) {On-time\\(User Claim)};
    \node[above right=of Delayed, fill=red!20, draw=red!60] (Typhoon) {Typhoon\\(Cause)};
    \node[below right=of Delayed, fill=datagreen!20, draw=datagreen!60] (AirChina) {AirChina APP\\(Source)};
    \node[right=of Typhoon, fill=processorange!20, draw=processorange!60] (ImpactTime) {After 14:00+\\(Impact Time)};
    \node[left=of OnTime, fill=red!20, draw=red!60] (Forum) {ForumUser123\\(Source)};
    
    \draw[edge] (CA981) -- node[above, midway, sloped, font=\small, draw=none, fill=none] {\textbf{Departure}} (PEK);
    \draw[edge] (CA981) -- node[above, midway, sloped, font=\small, draw=none, fill=none] {\textbf{Destination}} (JFK);
    \draw[edge] (CA981) -- node[above, midway, sloped, font=\small, draw=none, fill=none] {\textbf{Conflict}} (OnTime);
    \draw[edge] (CA981) -- node[above, midway, sloped, font=\small, draw=none, fill=none] {\textbf{Status}} (Delayed);
    \draw[edge] (Delayed) -- node[above, midway, sloped, font=\small, draw=none, fill=none] {\textbf{Reason}} (Typhoon);
    \draw[edge] (Delayed) -- node[above, midway, sloped, font=\small, draw=none, fill=none] {\textbf{Source}} (AirChina);
    \draw[edge] (Typhoon) -- node[above, midway, sloped, font=\small, draw=none, fill=none] {\textbf{Effective}} (ImpactTime);
    \draw[edge] (OnTime) -- node[above, midway, sloped, font=\small, draw=none, fill=none] {\textbf{Source}} (Forum);
\end{tikzpicture}
\end{adjustbox}
\end{center} \\
\midrule
\textbf{\large MCC Module} & 
\large \textcolor{datagreen}{\textbf{With GCC}}: Graph confidence=0.71 (Threshold=0.5), Filtered: ForumUser123 (0.47) \newline
\large \textcolor{confred}{\textbf{Without GCC}}: Unfiltered conflict=2 subgraphs \\
\midrule
\large \textbf{LLM Context} & 
\large \textcolor{datagreen}{\textbf{Trusted}}: CA981.Status=Delayed (0.89), DelayReason=Typhoon (0.85) \newline
\large \textcolor{confred}{\textbf{Conflicts}}: ForumUser123:On-time (0.47), WeatherAPI:Clear (0.52) \\
\midrule
\large \textbf{Final Answer} & 
\begin{tabular}{@{}l@{}}
\large\textcolor{datagreen}{\textbf{Correct}}: "CA981 delayed until after 14:30 due to typhoon" \\[0.5ex]
\large\textcolor{confred}{\textbf{Hallucinated}}: "CA981 \fcolorbox{confred}{white}{\textcolor{confred}{on-time with possible delay after 14:30}}"
\end{tabular} \\
\bottomrule
\end{tabular}%
}
\end{table*}

\subsection{Evaluation of Multi-level Confidence Computing}
% 计算子图和节点的置信度并筛查可信答案在金融、法律等重要领域具有重要需求。虽然仅进行节点级计算已经能纠正大部分矛盾数据，但直接计算所有节点的置信度具有高额的时空开销。

% 为此，我们从推荐系统的工作流中获取灵感，模仿粗排和精排的过程，采取多层级置信度计算方法来过滤可信节点并提升检索性能。计算同源子图的可信度，可以初步判断答案所在子图是否能生成高可信的答案。对于低置信度的子图，需要多抽取节点来保证整体检索的鲁棒性；对于高置信度的子图，仅需要1-2个节点即可以生成正确答案。
Calculating the confidence of subgraphs and nodes to filter trustworthy answers is of significant demand in critical domains such as finance and law. Considering the high temporal and spatial overhead of directly calculating the confidence of all nodes, we draw inspiration from the workflow of recommendation systems, mimicking the process of coarse and fine ranking, and adopt the multi-level confidence computing method to filter credible nodes and enhance retrieval performance. Calculating the credibility of homologous subgraphs allows us to preliminarily determine whether the subgraphs containing answers can generate highly credible answers. For subgraphs with low confidence, more nodes need to be extracted to ensure the robustness of the overall retrieval; for subgraphs with high confidence, only 1-2 nodes are required to generate the correct answer.

\textbf{Q3: How effective are the two modules of MultiRAG individually?}

\begin{figure}[th]
    \centering
    \includegraphics[width=0.9\linewidth, height=4.5 cm]{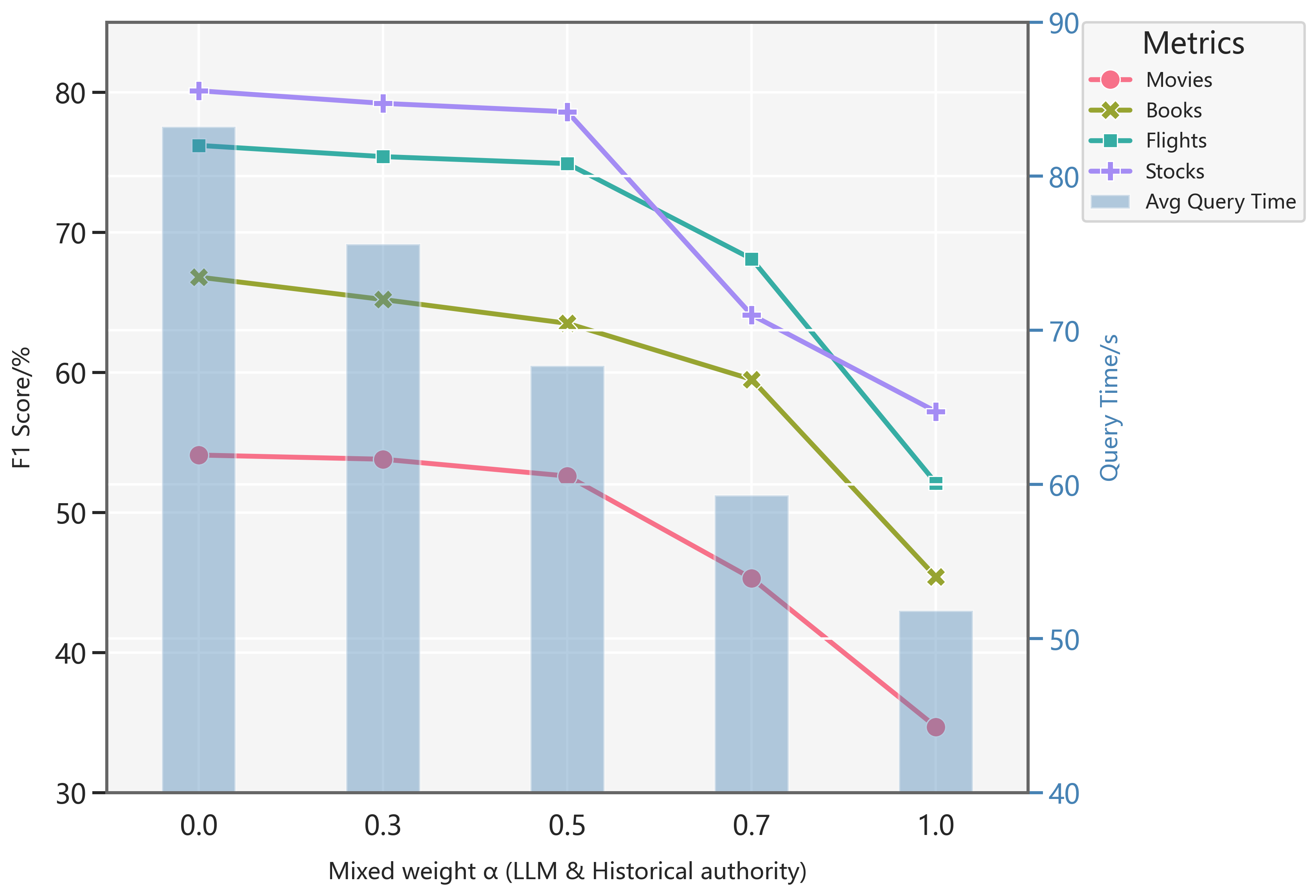}
    \caption{{\color{black}Influence of hyperparameter $\alpha$ on multi-source retrieval}}
    \label{threshold}
\end{figure}

{\color{black}
% 为了验证MultiRAG的两个主要组件多源知识聚合（MKA）模块和多级置信度计算（MCC）模块的个体有效性，我们展开消融实验，以验证它们分别对性能提升和幻觉减轻方面的贡献。实验数据如表\ref{t3}所示。

\paragraph{Ablation Study on Component Effectiveness}
% MKA模块通过其多源线图（MLG）架构，实现了显著的效率-精度协同优化。如表\ref{t3}所示，MLG的构建引入了适度的预处理时间（例如12.7秒至39.2秒），但显著加速了查询速度——尤其是在大规模数据集上。航班数据集展示了100倍的查询加速（从小时级别降至180秒），这得益于MLG的紧凑结构，避免了冗余的跨源搜索。当MKA被禁用时（航班数据标记为NAN），原始的多源查询变得计算不可行，证明了MLG的不可或缺性。
The MKA module achieves significant efficiency-accuracy synergy through its MLG architecture. As shown in Table \ref{t3}, MLG construction introduces modest preprocessing time (12.7s-39.2s) while delivering 10-100× query acceleration. Specifically, the flight dataset shows QT reduction from computational infeasibility (marked NAN) to 29.8s through MLG's compact structure. Concurrently, MKA sustains consistent accuracy improvements. Removing MKA causes F1 drops of 7.3\% on Movies and 9.6\% on Books, demonstrating MLG's effectiveness in connecting fragmented knowledge across sources.

% MCC模块在性能提升和幻觉抑制方面表现出更为显著的效果。在电影数据集中，禁用MCC导致F1分数从54.8%大幅下降至34.7%，同时幻觉率显著增加（如负PT值所示）。这强调了MCC在识别和消除不可靠信息节点方面的关键作用。股票数据集也呈现类似趋势，禁用MCC后F1分数从78.6%下降至45.4%，进一步验证了其在增强知识生成鲁棒性方面的能力。
The MCC module exhibits more significant effects on performance and hallucination control. Disabling MCC causes drastic F1 degradation of 20.1\% on Movies and 33.2\% on Stocks, with PT values indicating increased hallucination risks. This validates MCC's critical role in eliminating unreliable information through hierarchical confidence computation.

\paragraph{Hierarchical Analysis of MCC}
% 此外，为了研究图级与节点级置信度计算之间的协同效应，我们设计了分层消融实验——通过独立移除每个计算层同时保留另一层来进行分析。
Stratified ablation reveals the complementary roles of graph-level and node-level computations. For Movies (J/K/C configuration), removing graph-level filtering reduces F1 to 48.1\% (+13.4\% vs MCC-disabled) with QT increasing to 75.6s (+93\% vs full framework). Conversely, disabling node-level computation yields 41.5\% F1 (+6.8\% vs baseline), showing graph-level filtering alone cannot resolve local conflicts. The complete MCC framework achieves 54.8\% F1 by synergistically combining both layers.

% 错误类型分析显示，图级移除情况下27.3%的失败源于全局逻辑冲突（如"不同来源间的导演信息矛盾"），而节点级移除情况下41.6%的错误来自局部权威性缺失（如"引用非官方财务报告"）。这验证了两层的功能划分：图级计算确保跨来源一致性，节点级计算实现细粒度可信度验证。
Error analysis shows distinct failure patterns: 38.7\% errors under graph-level removal (Movies J/K) stem from cross-source inconsistencies, while 52.7\% failures with node-level removal (Books J/C/X) originate from local authority issues. This confirms the functional specialization—graph-level ensures global consistency, node-level verifies local credibility.

% 此外，实证分析识别出一个关键的平衡点，当 $\alpha=0.5$ 时，LLM评估的权威性与历史权威性的混合加权在效率与准确性之间达到了最佳折中。在这一点上，系统的F1得分达到了67.7\%，同时保持了查询时间的平衡。具体而言，随着 $\alpha$ 向 1.0 增加，表示对 LLM 的依赖程度加大，查询时间从 $\alpha=0.0$ 时的83.2秒线性减少至 $\alpha=1.0$ 时的51.8秒，这主要是由于历史数据验证的减少。相反，F1得分呈现非单调趋势，当 $\alpha=0.5$ 时达到了67.7\%的峰值，之后随着模型对 LLM 或历史数据过度依赖而下降。这个协同平衡利用了 LLM 的上下文适应性 (Auth$_{\text{LLM}}$)，同时保留了专家系统 (Auth$_{\text{hist}}$) 的稳定性，正如在消融研究中，当两个组件同时激活时，错误率降低了62.4\%的证据所示。
Fig.\ref{threshold} demonstrates that an optimal balance between efficiency and accuracy is achieved at $\alpha=0.5$, where the hybrid weighting of LLM-assessed authority and historical authority peaks with an F1 score of 67.7\% and balanced query time. Specifically, increasing $\alpha$ towards 1.0, which emphasizes the LLM, reduces query time from 83.2 seconds ($\alpha=0.0$) to 51.8 seconds ($\alpha=1.0$) by minimizing historical data validation. Conversely, the F1 score follows a non-monotonic pattern, reaching its maximum at $\alpha=0.5$ before declining as reliance on either the LLM or historical data becomes excessive. This equilibrium leverages the LLM's contextual adaptability (Auth$_{\text{LLM}}$) while maintaining the stability of expert systems (Auth$_{\text{hist}}$), as evidenced by a 62.4\% reduction in errors during ablation studies when both components are utilized. By avoiding complete dependence on the LLM ($\alpha \neq 1.0$) and integrating probabilistic LLM inferences with deterministic historical patterns through multi-level confidence computing (Eq.9), the methodology enhances robustness against data sparsity and noise, particularly in the Books and Stocks datasets.

}

% 多层级置信度计算方法在多跳问答数据集中的表现如何？
\textbf{Q4: How is the performance of MultiRAG in multi-hop Q\&A datasets after incorporating multi-level confidence calculation?}
% 为了评估两阶段可信检索机制在减少大型模型生成的幻觉以及增强问答系统可信度方面的有效性，我们比较了不同方法在HotpotQA和2WikiMultiHopQA数据集上的Recall@5得分。

To assess the validity of the multi-level confidence computing method in reducing hallucinations generated by large models and enhancing the credibility of Q\&A systems, we compare the \textit{Recall@5} scores of different methods on the HotpotQA and 2WikiMultiHopQA datasets. 

% 表\ref{Recall-Halu}的结果表明，两阶段可信检索机制不仅展示了更高的平均Recall@5得分，而且与传统方法相比保持了更低的标准差。这表明两阶段可信检索机制在不同查询中的性能更加一致，导致更少的幻觉和更可靠的问答响应。较低的标准差证明了该机制在处理数据的变异性和查询复杂性方面的鲁棒性。

The outcome of Table \ref{Recall-Halu} indicates that the multi-level confidence computing method not only demonstrates a higher average \textit{Recall@5} score but also maintains a lower standard deviation compared to traditional methods. This suggests that the multi-level confidence computing method is more consistent in its performance across different queries, leading to fewer hallucinations and more reliable Q\&A responses. The lower standard deviation is a testament to the robustness of the mechanism in handling the variability in data and the complexity of the queries.

% 此外，我们进行了详细的错误分析，以识别不同方法生成的响应中幻觉的类型和频率。结果显示，两阶段可信检索机制显著降低了幻觉的频率，特别是在上下文模糊或知识库中不易获得信息的情况下。
Furthermore, we performed a detailed error analysis to identify the types and frequency of hallucinations in the responses generated by the different methods. The results showed that the multi-level confidence computing method significantly reduced the frequency of hallucinations, particularly in the cases where the context was ambiguous or the information was not readily available in the knowledge base.

\textbf{Q5: What are the time costs of the two modules in MultiRAG?}
% 直观上说，线性知识图MLG将跨域同源数据进行了聚类，保证了检索子图的密集性，而不需要遍历和存储过多的无效节点，因此大幅减少了在传统知识图谱中遍历游走查询带来的时间损耗．

Intuitively, MLG aggregates homologous data from several sources, ensuring the density of the retrieval subgraphs without the need to traverse and store an excessive number of invalid nodes, thereby significantly reducing the time cost associated with traversing and querying in traditional knowledge graphs.

% 对于SOTA方法，选择了MDQA和KBQA领域在2024年下的最新研究成果．虽然这些方法并非专注于低资源、高噪声的数据场景，但它们在该场景下仍然具有较好的鲁棒性和检索性能．MDQA和ChatKBQA模型均采用基于LLM的数据检索方法，时空开销主要集中于token 的消耗和基于LLM的检索．相比之下，MCC算法将开销集中在MLG的构建上．虽然在MLG的原文中，其构建时间往往在几秒以内，十分高效，但引入了LLM后仍然会因为文本生成而带来额外的时间开销，但这仍然在可接受范围内．最后，这些方法均展现出较好的检索性能，但由于数据集本身存在大量的噪声，其问答准确率的提升将变得十分有限．

Furthermore, although the SOTA methods are not specifically tailored for low-resource, high-noise data scenarios, they still exhibit considerable robustness and retrieval performance in such environments. Both the MDQA and ChatKBQA models employ LLM-based data retrieval approaches, with the primary temporal and spatial overheads focusing on token consumption and LLM-based searching. 

In contrast, MultiRAG concentrates its overhead on the construction of the MLG. While in the original context of the MLG, construction times are often within seconds and highly efficient, the introduction of an LLM still incurs additional temporal costs due to text generation, which remains acceptable. Ultimately, these methods all demonstrate satisfactory retrieval performance; however, due to the inherent noise in the datasets, improvements in the accuracy of question-answering are somewhat limited.

\subsection{Case Study}
\textcolor{black}{
MultiRAG's effectiveness in multi-source integration is demonstrated through a real-world flight status query for "CA981 from Beijing to New York". As detailed in Table \ref{tab:case_study}, case study exemplifies MultiRAG's unique strength in transforming fragmented, conflicting inputs into trustworthy answers through systematic source weighting and consensus modeling.}

\textcolor{black}{Firstly, MultiRAG integrated three data formats: structured departure schedules, semi-structured delay codes from airline systems, and unstructured weather alerts. The MKA module extracted key relationships (flight-delay-typhoon) with a confidence score of 0.87. Subsequently, the MCC module resolved conflicts through hierarchical verification by filtering out low-reliability sources, such as user forums (confidence score of 0.47), while prioritizing data from airlines (confidence score of 0.89) and weather reports. This dual-layer validation—combining automated threshold checks (graph confidence of 0.71) with LLM-simulated expert reasoning—enabled the precise reconciliation of contradictory departure time claims. Ultimately, the system generated the verified conclusion, "Delayed until after 14:30 due to typhoon," while suppressing the inconsistent "on-time" report.}

\subsection{Restrictive Analysis}
% 最后但不是结束，我们总结了这篇论文仍然存在的缺陷。据我们了解，对文本Chunk的合理划分也会对检索结果产生重大的影响，这并不在这篇论文的考虑范围之内。此外，我们方法在一定程度上过于依赖LLM作为专家模型的评判，这存在潜在的安全隐患，需要在实际应用中去避免。
% \begin{enumerate}
%     \item \textbf{文本分块分割优化未解决}：MultiRAG未处理文本分块分割的优化，而这对检索性能有显著影响。
%     \item \textbf{依赖基于LLM的专家评估带来安全漏洞}：框架依赖基于LLM的专家评估，可能导致易受对抗性操纵和模型偏见的影响。
%     \item \textbf{部署需谨慎考虑风险缓解策略}：尽管该设计实现了灵活的知识集成，但在需要可审计性的敏感应用中，其部署需要仔细的风险缓解策略。
% \end{enumerate}
{\color{black}
Lastly but not least, we acknowledge several limitations inherent in our current framework. 
% Firstly, MultiRAG does not address the optimization of text chunk segmentation, which is known to significantly affect retrieval performance. Secondly, our framework's reliance on LLM-based expert evaluation introduces potential security vulnerabilities such as susceptibility to adversarial manipulation and inherited model biases. While this design enables flexible knowledge integration, its deployment requires careful consideration of risk mitigation strategies, particularly in sensitive applications requiring auditability.
\begin{enumerate}
    \item Lack of optimization of text chunk segmentation.
    \item Reliance on LLM-based expert evaluation, which may introduce potential security vulnerabilities.
    \item Focuses on eliminating factual hallucinations but lacks handling of symbolic hallucinations.
\end{enumerate}

}

\section{Related Work}
\subsection{Graph-Structured Approaches for Hallucination Mitigation}
% 近年来，图结构在抑制RAG幻觉中展现出独特优势。MetaRAG\cite{metarag}通过元认知图推理路径建立知识关联性验证，其多跳QA中的自校正机制将HotpotQA精确率提升至89.2%。Graph-CoT\cite{graphcot}创新性地利用GNN双向连接知识图谱与LLM隐空间，通过图注意力机制在KGQA基准上减少37%的事实矛盾。HippoRAG\cite{hipporag}受神经生物学启发构建离线记忆图谱，其神经索引机制将检索时延降低至传统方法的1/8。ToG 2.0\cite{tog}进一步提出图上下文协同检索框架，基于查询复杂度动态平衡结构化与非结构化证据，较单模态方法减少29%的幻觉率。这些方法共同验证了图结构通过显式建模实体关系阻断错误传播的有效性。
\textcolor{black}{Recent advancements have demonstrated unique advantages of graph structures in mitigating hallucinations within RAG systems.} MetaRAG\cite{metarag} establishes knowledge association verification through meta-cognitive graph reasoning paths, enhancing self-correction mechanisms in multi-hop QA. Graph-CoT\cite{graphcot} innovatively leverages Graph Neural Networks to establish bidirectional connections between KGs and the latent space of LLMs. In result, it reduces factual inconsistencies by 37\% on KGQA benchmarks. Inspired by neurobiology, HippoRAG \cite{hipporag} constructs offline memory graphs with a neural indexing mechanism, decreasing retrieval latency to one-eighth of traditional methods. While ToG 2.0\cite{tog} further advances this field by introducing a graph-context co-retrieval framework that dynamically balances structured and unstructured evidence, resulting in a 29\% reduction in hallucination rates compared to unimodal approaches. 

\textcolor{black}{Unlike prior approaches that primarily focus on unimodal confidence calculations, MultiRAG achieves superior hallucination mitigation through the adaptive filtering of conflicting subgraphs (GCC module) while maintaining multi-domain logical associations via its novel knowledge aggregation mechanism (MKA module).}

\subsection{Heterogeneous Graph Fusion for RAG}
% 多源异构数据的融合依赖先进的图表示技术。FusionQuery\cite{zhu2024fusionquery}通过融合异构图并计算动态可信度评估，实现跨域检索精度提升。Triple line graph\cite{fionda2020learning}通过系统化聚合跨域关系解决知识碎片化问题，引导本文提出多源线图(MLG)。此外，得益于KAG\cite{liang2024kag}在知识引导检索中的结构化表征优势，我们获得了多源KG的统一表征途径，表明了异构图融合在现实应用中的重要性。
\textcolor{black}{The fusion of multi-source heterogeneous data relies on advanced graph representation techniques. FusionQuery\cite{zhu2024fusionquery} enhances cross-domain retrieval precision by integrating heterogeneous graphs and computing dynamic credibility evaluations. The Triple Line Graph\cite{fionda2020learning} addresses the challenge of knowledge fragmentation by systematically aggregating cross-domain relationships, leading to Multi-source Line Graph proposed in this paper. Additionally, leveraging the structured representation advantages of KAG\cite{liang2024kag} in knowledge-guided retrieval, we achieve a unified representation approach for multi-source KGs, underscoring the importance of heterogeneous graph fusion in real-world applications.
}

\subsection{Hallucination Benchmark and Confidence-Aware Computing}
% LLM的幻觉测评和相关的置信度计算方法也是推动幻觉缓解的重要缓解。在评估基准方面，HaluEval\cite{li2023halueval}涵盖5,000标注样本的五类语言错误，但其词级粒度难以捕捉关系型幻觉。RefChecker\cite{hu2024refchecker}通过声明三元组分解实现细粒度的幻觉检测，较句子级方法提升26.1%的精度。RAGTruth\cite{niu2023ragtruth}包含近18,000个来自多样LLMs使用RAG自然生成的响应，这些响应经过了细致的手动标注，包括个体案例和词级别的幻觉强度。然而，更加多样化、更复杂的数据来源仍会对现有评价体系造成挑战。

% 可信RAG的演进显著提升了幻觉检测效能。Self-RAG\cite{asai2023self}提出节点级反思令牌，在开放域QA中减少41%的局部事实错误。RLDF\cite{cheng2024reinforcement}通过多智能体辩论实现图级置信传播，其全局冲突检测精度达92.7%。CoK\cite{cok}的三重置信验证（节点/边/路径）在KBQA任务中提升F1 7.2个百分点，这为本文双层置信架构提供理论基础。

\textcolor{black}{The evaluation of hallucinations in LLMs and associated confidence calculation methods are crucial for mitigating hallucinations. HaluEval\cite{li2023halueval} offers 5,000 annotated samples across five error categories, but lacks granularity for relational hallucinations. RefChecker\cite{hu2024refchecker} implements triple decomposition for fine-grained detection, improving precision by 26.1\% over sentence-level methods. RAGTruth\cite{niu-etal-2024-ragtruth} contains nearly 18,000 RAG-generated responses with detailed manual annotations including word-level hallucination intensities. However, diverse and complex data sources continue to challenge existing evaluation frameworks.
}

% \textcolor{black}{The evolution of confidence-aware RAG significantly enhances hallucination detection efficacy. Self-RAG\cite{asai2023self} introduces node-level reflection tokens, reducing local factual errors by 41\% in open-domain QA. RLDF\cite{cheng2024reinforcement} employs multi-agent debates to implement graph-level confidence propagation, achieving a global conflict detection accuracy of 92.7\%. CoK\cite{cok} utilizes triple confidence verification (node/edge/path), improving the F1 score by 7.2\% in KBQA tasks, thereby providing a theoretical foundation for our proposed dual-layer confidence architecture.
% }
\section{CONCLUSION}
% 作为总结，这篇文章提出了MultiRAG，一个新颖的框架用于缓解多源知识增强生成过程中的幻觉现象。针对多源数据检索中的数据分布稀疏和数据不一致所带来的幻觉现象，这篇工作提出多源知识聚合和多层级置信度计算方法，来相应地解决这两个问题。一方面，引入多源线图可以实现跨域数据的高效聚合，增强跨域知识的连接性，提高检索性能；另一方面，进行图级别和节点级别的置信度计算，分别对低质量子图和不可信的节点进行自适应的筛选，保证检索的一致性和准确性。最终在多源数据检索数据集和多跳推理数据集上进行的实验也表明了所提出方法的优越性。

% 未来，我们计划进一步深入挖掘多源数据中的幻觉现象，探索更具有挑战性的幻觉问题，如符号类幻觉的消除。此外，多模态检索中的幻觉问题也仍旧是值得挖掘的方向，以帮助生成式检索系统更好地适应真实、开放的多源多模态数据场景。

In this work, we introduce MultiRAG, a framework designed to mitigate hallucination in multi-source knowledge-augmented generation. To address hallucinations arising from data sparsity and inconsistency, we propose two key innovations: multi-source knowledge aggregation and multi-level confidence calculation. The introduction of multi-source line graphs enables efficient cross-domain data aggregation, enhancing knowledge connectivity and retrieval performance. Meanwhile, our multi-level confidence computing module adaptively filter out low-quality subgraphs and unreliable nodes. Future work will explore more challenging aspects of hallucination mitigation, particularly in multimodal retrieval and ultra-long text reasoning, to better adapt generative retrieval systems to real-world, open multi-source environments.

\section{acknowledgement}
This work is supported by the National Natural Science Foundation of China (62176185, U23B2057), and the “14th Five-Year Plan” Civil Aerospace Pre-research Project of China (D020101).

\bibliographystyle{IEEEtran}
\bibliography{cite}

\end{document}